\documentclass[twocolumn,showpacs,preprintnumbers,amsmath,amssymb,prb]{revtex4}
\usepackage{rotating} 
\usepackage{epsfig} 
\usepackage{amsfonts}
\usepackage{latexsym}
\usepackage{enumerate}
\usepackage{longtable}

\begin{document}

\title{\bf Systematic Mapping of the Hubbard Model to the 
Generalized $t$-$J$ Model}

\author{Alexander Reischl, Erwin M\"uller-Hartmann, and G\"otz
  S. Uhrig}

\affiliation{Institut f\"ur Theoretische Physik, Universit\"at zu
  K\"oln, Z\"ulpicher Str.~77, D-50937 K\"oln, Germany}

\date{\today} 

\begin{abstract}
The generalized $t$-$J$ model conserving the number of double occupancies is
constructed from the Hubbard model at and in the vicinity of half-filling at 
strong coupling. The construction is realized by a self-similar continuous
unitary transformation. The flow equation is closed by a truncation scheme
based on the spatial range of processes. We analyze the conditions under
which the $t$-$J$ model can be set up and we find that it can only
be defined for sufficiently large interaction. There, the parameters of the
effective model are determined.
\end{abstract}


\maketitle

\section{Introduction}
The Hubbard model serves as a prototype model for the description of
strongly correlated electron systems. It consists of electrons with
spin moving on a lattice. They repel each other on-site which leads to a
complex interplay of magnetic and charge degrees of 
freedom. At half-filling, the model displays insulating behavior in the
limit of strong interaction. In this limit, the model
can be mapped onto a pure spin Heisenberg model
\cite{ander59,klein73,takah77,harri67,macdo88,stein97}. On the other side,
vanishing interaction makes it a 
metal. In-between an intricate metal-insulator transition (MIT) takes
place which is well understood only in extreme cases. In one
dimension, the Bethe ansatz tells us that
arbitrarily small interactions render the model
insulating \cite{lieb68}. In infinite dimensions, the dynamical 
mean-field theory allows to make statements about the value of the critical
interaction \cite{gebha97,georg96,eastw03} if  long-range magnetic
order is suppressed. It is argued that this suppression can be achieved
physically by introducing frustration.

It is the aim of the present paper to provide a systematic and controlled
non-perturbative derivation of the $t$-$J$ model from the Hubbard model. 
Most importantly, we will discuss under which circumstances the reduction of 
the Hubbard model to a $t$-$J$ model is justified.
The $t$-$J$ model is the effective model describing
fermions which interact and hop \emph{without} creating  or
annihilating double occupancies (DOs). This means that never two electrons
occur on the same site.  Thus the reduction to a $t$-$J$ model corresponds
to the elimination of charge fluctuations.
The parameters of the effective model will
be computed quantitatively. Our results reach in two ways beyond
what has been done before
\cite{ander59,harri67,klein73,takah77,macdo88,stein97}. 
First, we provide non-perturbative results.  Second, we discuss
the matrix elements of hole motion and interaction in a 
systematic way beyond the zeroth order. Both ingredients help us to 
discuss the breakdown of the reduction to a $t$-$J$ model.

The reference ensemble in our approach is the magnetically 
completely disordered half-filled model (see Sect.\ III). This constitutes the
vacuum in our calculations. In this sense, we
start from the Hubbard model \emph{at} half-filling.
But our approach naturally  keeps also track of
the dynamics of holes. This is unavoidable since we have to know
how virtual intermediate states evolve and  they contain generically 
charge excitations. The parts of the effective model which 
describe the dynamics of holes apply also to holes inserted externally
by doping. In this sense, the effective model describes also the physics of
a vicinity of half-filling. But it is beyond the scope of the present paper
to discuss the doping dependence of the \emph{couplings} of the effective
model since this would correspond to a change of the reference ensemble.

\begin{figure}[htb]
  \begin{center}
    \includegraphics[width=\columnwidth]{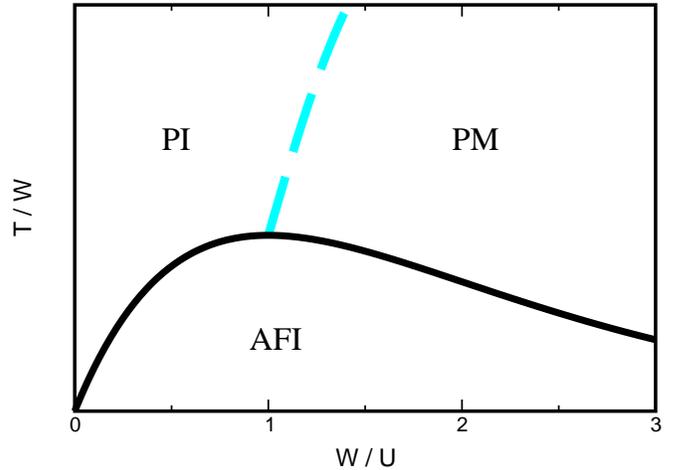}
    \caption{Schematic phase diagram of the Hubbard model at half-filling
      in higher dimensions $d>2$. The tick labels at the $x$-axis are
      indicative only. The bare band width is denoted by $W$, the 
      Hubbard repulsion by $U$, the temperature by $T$; 
      PI stands for paramagnetic insulator, PM for paramagnetic metal and AFI 
      for antiferromagnetic insulator. The dashed line indicates the
      transition or the crossover from insulator to metal.}
    \label{fig:intro}
  \end{center}
\end{figure}

In Fig.~\ref{fig:intro}
a schematic phase diagram is shown which summarizes the currently assumed
picture. This picture is motivated from various mean-field computations,
in particular in infinite dimensions \cite{georg96}.
 For large repulsion $U$ (left side in the figure), 
the electrons are localized so that the system is insulating. The spins
interact via the magnetic exchange $J$ of the order of $4t^2/U$ which
sets the energy scale for the transition between a long-range magnetically
ordered phase (AFI) and a disordered phase (PI). For decreasing $U$,
the ground state remains insulating unless the magnetic order is 
fully suppressed. If the spin state becomes sufficiently disordered,
for instance at higher temperatures, the system becomes conducting (PM).
The change from paramagnetic insulator to paramagnetic metal is a
first order transition in infinite dimensions and for not too large 
temperatures\cite{georg96}. 
In general, it is to be expected that the system undergoes a crossover.
In two dimensions in particular, the case investigated here,
no long-range antiferromagnetism occurs at finite temperatures \cite{mermi66} 
so that the AFI collapses and only the disordered phases occur.

Once the interaction is such that the system is a paramagnetic metal,
charge excitations are possible at zero energy. Hence, it cannot be
expected that the elimination of charge fluctuations, 
virtual double occupancies, is possible. 
This means that the reduction
of the Hubbard model to a $t$-$J$ model can only be defined as long as the
paramagnetic phase is insulating. The sole occurence of an 
insulating phase with long-range order is not sufficient. The insulating
behavior in the long-range ordered AFI phase for interactions where the 
paramagnetic phase is metallic is due to the concomitant 
breaking of the translation symmetry. Hence it resembles rather a band
insulator than a genuine Mott-Hubbard insulator \cite{notiz1}.
The $t$-$J$ model, however, describes low-lying degrees of freedom,
namely the degrees of freedom of the spins and of doped charge carriers,
without the precondition of a certain long-range order. The derivation
of the $t$-$J$ model requires to eliminate the virtual charge fluctuations
while leaving the state of the spins and of the
doped charges unspecified. One may consider the unspecified low-lying
degrees of freedom to be at infinite temperature in a completely disordered
mixture. The phase diagram in Fig.~\ref{fig:intro}
tells us that there must be a certain $U$ of the order of the band width
$W$ below which the Hubbard model cannot be represented by a $t$-$J$ model.

The method of continuous unitary transformations (CUTs) has been
proposed by Wegner in the context of condensed matter physics \cite{wegne94}.
Simultaneously, similar approaches were tested by G{\l}azek and Wilson
in high energy physics \cite{glaze93,glaze94}. 
The proper choice of an infinitesimal generator brings the many-body
problem under study into a more tractable form by decoupling different
sectors of the Hamiltonian. In many cases it is possible to advance to
an effective model that is tractable by some other means
while retaining the complex physics of the original problem 
\cite{kehre94,stein97,knett00a,heidb02a,heidb02b,white02a}.
We will apply a CUT to the Hubbard model in order to eliminate
operators that couple states of different double occupancy. 
This has been done before by Stein \cite{stein97} in a perturbative
fashion. In contrast to Stein's perturbative treatment we will adopt a
self-similar truncation scheme for the 
operators emerging during the transformation.
This self-similar CUT retains terms of a certain structure.

The paper is organized as follows. The Hubbard model is introduced in
Sect.~\ref{sec:hub}. Sect.~\ref{sec:cut} explains the method of
self-similar CUTs.  A transparent example is given to
illustrate  technical details and characteristic properties of the CUT
applied to the Hubbard model. The results for the range of validity of
the mapping and for the effective parameters are presented in 
Sect.~\ref{sec:res}. Evidence is provided that the mapping cannot be 
defined for all ratios $W/U$. The physical implications of our findings are 
discussed in Sect.~\ref{sec:dis}. 
A brief summary is given finally in Sect.~\ref{sec:sum}.

\section{Hubbard model and Operators}\label{sec:hub}
Our starting point is the Hubbard model on a 
two-dimensional square lattice in the vicinity of
 half-filling with nearest-neighbor hopping. 
Since the calculation covers also the motion and
interaction of charge carriers it is not limited to
strictly half-filling but covers also small hole concentrations.

The Hamilton operator is split into a kinetic part
${H}_t$ and an interaction ${H}_U$
\begin{subequations}
\label{hamilton}
\begin{eqnarray}
  {H} &=& {H}_U+ {H}_{t} \\ 
  {H}_U &=& U \sum_{i} (
  n_{i,\uparrow}-1/2)(n_{i,\downarrow}-1/2)\\
  &=&(U/2)\left(\hat D-N/2\right) \nonumber \\ 
  {H}_{t}&=& t \sum_{\langle i,j\rangle, \sigma} ( c^\dagger_{i,\sigma}
  c^{\phantom{\dagger}}_{j,\sigma}+\text{h.c.}),
\end{eqnarray}
\end{subequations}
where $N$ is the number of sites, $c^\dagger_{i,\sigma}$,
$c^{\phantom{\dagger}}_{i,\sigma}$ are creation and 
annihilation operators of an electron on site $i$ with spin $\sigma
\in \{\uparrow, \downarrow\}$ and
$n_{i,\sigma}=c^\dagger_{i,\sigma}c^{\phantom{\dagger}}_{i,\sigma}$ 
is the number operator. Hopping is only possible between nearest neighbor 
(NN) sites as indicated by $\langle i,j\rangle$ in the sum in $H_t$.
The bare hopping coefficient
is given by $t$ so that in two dimensions the band width
is $W=8t$. It turns out to be convenient to denote all results in terms of
the band width $W$. 
 
We also define the operator 
$\hat D :=\sum_i [n_{i,\uparrow}n_{i,\downarrow} + 
(1-n_{i,\uparrow})(1-n_{i,\downarrow})]$ that counts the number of double 
occupancies  (DOs) of particles {\it and} holes.  ${H}_U$ describes the 
repulsion of electrons of opposite spin on the same site. For large $U$ the 
density of states splits into an upper and a lower Hubbard band separated by
an energy of order $U$.  In the half-filled case, in the limit of
infinite $U$ the electrons are fixed on their lattice site by the
constraint of having no DO.  Reducing $U$ from this limit the 
electrons are gradually allowed to move producing DOs intermediately.
But the physics remains still rather local.  In the following, we want to
map the Hubbard model for finite $U$ in a systematic way onto an effective 
model that conserves the number of DOs.  
The guiding idea will be that the physics
can be incorporated into an effective model that contains operators
that describe rather local processes. For too small values of $U$ the 
vanishing of the charge gap  will make the mapping impossible.

\section{Continuous Unitary Transformation}\label{sec:cut}
In this section, the technical details of our calculations are presented.
The method of continuous unitary transformations (CUT) introduces a
continuous auxiliary variable $\ell$ and transforms the Hamiltonian
according to the flow equation \cite{wegne94}
\begin{equation}
  \label{fleq}
  \frac{d}{d\ell}{H}(\ell)=[\eta(\ell),{H}(\ell)]
\end{equation}
with an antihermitian generator $\eta(\ell)$. The operators ${H}(\ell)$ for
different values of $\ell$ are unitarily equivalent.  The transformation starts
with the initial condition ${H}(\ell=0)={H}$ and ends with an
effective Hamiltonian 
at $\ell \to \infty$. We want the effective Hamiltonian to conserve
the number of DOs and therefore use the generator 
\cite{stein97}
\begin{equation}
  \label{eta}
  \eta(\ell)= [ \hat D , H(\ell) ]\ .
\end{equation}
Except for an overall factor\cite{notiz4}, the above $\eta$ coincides with 
$\eta$ in Refs.~\onlinecite{mielk98,knett00a}, where it was defined
using the {\em sign} of the change in the number of quasiparticles
ensuring that the block-band structure of the
Hamiltonian is conserved. This means that  during the flow  no
operators are generated that change the number of DOs
by a value other than $0$, $-2$ and $2$.

To see how the CUT works, it is helpful to classify the operators in 
the kinetic part of
the hamiltonian according to their effect on the number of DOs
\begin{subequations}
\label{eq:ham}
\begin{eqnarray}
  {H}_t &=& T_0 + T_{+2} + T_{-2} \\ 
T_0&=& t_0\sum_{\langle i,j\rangle, \sigma}
  \Big[(1-n^{\phantom{\dagger}}_{i,\sigma}) c^\dagger_{i,\overline{\sigma}}
  c^{\phantom{\dagger}}_{j,\overline{\sigma}} 
(1-n^{\phantom{\dagger}}_{j,\sigma})\nonumber \\
\label{eq:T0}
&&\phantom{t_0\sum_{{ \langle i,j\rangle, \sigma}}\Big[}
  +n^{\phantom{\dagger}}_{i,\sigma} c^\dagger_{i,\overline{\sigma}}
   c^{\phantom{\dagger}}_{j,\overline{\sigma}} 
n^{\phantom{\dagger}}_{j,\sigma}+\text{h.c.}\Big]\\ 
  T_{+2} &=& t_{+2}\sum_{\langle i,j\rangle,\sigma}
  \Big[ n^{\phantom{\dagger}}_{i,\sigma} c^\dagger_{i,\overline{\sigma}}
  c^{\phantom{\dagger}}_{j,\overline{\sigma}} 
(1-n^{\phantom{\dagger}}_{j,\sigma})\nonumber \\
&&\phantom{t_{+2} \sum_{{ \langle i,j\rangle, \sigma}}\Big[}
+ n^{\phantom{\dagger}}_{j,\sigma} c^\dagger_{j,\overline{\sigma}}
  c^{\phantom{\dagger}}_{i,\overline{\sigma}} 
(1-n^{\phantom{\dagger}}_{i,\sigma}) \Big] \\ 
  T_{-2} &=& t_{-2} \sum_{\langle i,j\rangle,\sigma}
  \Big[ (1-n^{\phantom{\dagger}}_{i,\sigma}) 
c^\dagger_{i,\overline{\sigma}}
  c^{\phantom{\dagger}}_{j,\overline{\sigma}} 
n^{\phantom{\dagger}}_{j,\sigma} \nonumber \\
&&\phantom{t_{-2}\sum_{{\langle i,j\rangle, \sigma}}\Big[} +
(1-n^{\phantom{\dagger}}_{j,\sigma}) c^\dagger_{j,\overline{\sigma}}
  c^{\phantom{\dagger}}_{i,\overline{\sigma}} 
n^{\phantom{\dagger}}_{i,\sigma} \Big]\ ,
\end{eqnarray}
\end{subequations}
where  $\overline{\sigma}=-\sigma$.
The projection operators ensure that $T_n$ changes the number of DOs by
$n$. 

Operators will be denoted in a standardized normal-ordered real
space representation (see Table \ref{tab:locop}). The pre\-fac\-tors of such a
product of local operators will be called  \emph{coefficient} of this
term in the Hamiltonian. 
Eq.\ (\ref{fleq}) will yield differential equations for these 
coefficients. The coefficients $t_0$, $t_{+2}$ and $t_{-2}$ coincide
 and are equal to the bare hopping $t$ in the beginning of the transformation. 
But they will develop differently during the flow. 

The operator $T_n$ yields a contribution $[\hat D,T_n(\ell)]=n T_n(\ell) $ to
$\eta$ leading to
\begin{equation}
  \frac{d}{d\ell}H= [\eta, H_U] + ... = - \frac{U}{2} n^2 T_n(\ell) + \ldots
\end{equation}
and thus to a suppression of any term not conserving the number
of DOs. 

In general the prescription (\ref{fleq}) will not produce closed
equations. One has to 
find a way to deal with the proliferating number of terms.
Stein \cite{stein97} set up a {\it perturbative} CUT for the Hubbard model
with the same generator (\ref{eta}). He ordered the terms produced
on the right hand side of (\ref{fleq}) according to
the power of $t/U$ with which they were generated.  The terms were
kept in a form consisting of long products of $T_0$, $T_{+2}$ and $T_{-2}$.

Contrary to this perturbative strategy, we will not keep the
commutators of $T_0$, $T_{+2}$ and $T_{-2}$ in the unevaluated form.
We compute the right hand side of the flow equation (\ref{fleq})
explicitly. This will produce terms already present in the Hamiltonian
and new terms. According to systematic rules the new terms
will be kept or discarded. Finally, a closed set of terms is reached.
The differential equations representing the flow equations of the
CUT are obtained by comparing the coefficients of all the retained terms.
Since both $\eta$ and
${H}$ are linear in the coefficients the right hand sides of the
differential equations are bilinear in the coefficients in close
analogy to conventional renormalization group equations.
 
To find a scheme to truncate the generation of new terms we 
make the following considerations. In the limit of large $U$ the motion
of the electrons at half-filling is suppressed by the energy cost of
DOs. Therefore the operators that become important first when going to
lower $U$ are those that describe local processes. To use this for the
truncation scheme we have to define a measure for the locality of a term. 
An undispensable prerequisite is a physically meaningful
and in particular \emph{unique} way to denote the operators.
Hence we define a sort of normal-ordering of the operator 
products\cite{notiz3}.
We will normal-order all terms with respect to the
half-filled paramagnetic reference ensemble represented by
the statistical operator
\begin{equation}
\label{refensemble0}
\hat \rho_0 := \prod_i (1/2)(|\!\uparrow\rangle_i {}_i\langle \uparrow\!|
+ |\!\downarrow\rangle_i {}_i\langle \downarrow\!|)
\end{equation}
where we use a basis of local states at each site $i$. 
There are four local states:  the empty site $|0\rangle$, two singly
occupied states with spin up und down, $|\!\uparrow\rangle$ and
$|\!\downarrow\rangle$, and the state with two electrons,
$|\!\uparrow\downarrow\rangle$.
The reference ensemble (\ref{refensemble0}) contains only
states with one electron per site since it is
half-filled. The up-spin and down-spin states are weighted equally 
to represent a  completely disordered paramagnetic reference ensemble.
Any deviation from the reference ensemble represents an excitation.

A product of $n$ local operators 
${\mathcal O}_{1} \cdot ... \cdot {\mathcal O}_{n}$ is normal-ordered if
any expectation value
\begin{subequations}
 \label{refensemble}
\begin{eqnarray}
\langle  {\mathcal O}_i \rangle_{\rm ref} &:=& 
\text{Tr} ({\mathcal O}_i\hat \rho_0)\\
&=& (1/2)({}_i\langle \uparrow\!|{\mathcal O}_i 
|\!\uparrow\rangle_i+{}_i\langle \downarrow\!|{\mathcal O}_i 
|\!\downarrow\rangle_i)
\end{eqnarray}
\end{subequations}
with the reference ensemble vanishes. In Table \ref{tab:locop} all local
operators are given in their normal-ordered form. Apart from the unity matrix 
$\openone$ all local operators in Tab.~\ref{tab:locop} are chosen such that 
the 
expectation value (\ref{refensemble}) with the reference ensemble yields zero. 
The operator 
$\bar{n}=n^{\phantom{\dagger}}_\uparrow+
n^{\phantom{\dagger}}_\downarrow-\openone$ counts
the number of electrons relative to half-filling. The Hubbard operator
$2n^{\phantom{\dagger}}_\uparrow 
n^{\phantom{\dagger}}_\downarrow- \bar{n}$ counts the number of
DOs. The empty state $|0\rangle$ and the doubly occupied state
$|\!\uparrow\downarrow\rangle$ yield 
$\langle 0|2n^{\phantom{\dagger}}_\uparrow n^{\phantom{\dagger}}_\downarrow- 
\bar{n}|0\rangle=
\langle \uparrow\downarrow\!|2n^{\phantom{\dagger}}_\uparrow 
n^{\phantom{\dagger}}_\downarrow- \bar{n}
|\!\uparrow\downarrow\rangle=1$ whereas expectation values with 
$|\!\uparrow\rangle$ and $|\!\downarrow\rangle$ vanish. For the spin operator 
$\sigma^z$ the sum (\ref{refensemble}) is zero as well. All other operators 
in Table \ref{tab:locop} are non-diagonal in the local basis and thus 
normal-ordered with respect to the reference ensemble.

\begin{table}[htb]
\begin{tabular}{|c|c|c|c|}
  \hline
  $ \openone $ &
  $\bar{n}=n^{\phantom{\dagger}}_\uparrow+
  n^{\phantom{\dagger}}_\downarrow-\openone$& 
  $\sigma^+=c^\dagger_\uparrow c^{\phantom{\dagger}}_\downarrow $& 
  $c^{\phantom{\dagger}}_\downarrow c^{\phantom{\dagger}}_\uparrow$ \\ 
%
  $\sigma^z=n^{\phantom{\dagger}}_\uparrow-
 n^{\phantom{\dagger}}_\downarrow$ & 
  $2n^{\phantom{\dagger}}_\uparrow n^{\phantom{\dagger}}_\downarrow-
 \bar{n}$&
  $\sigma^-=c^\dagger_\downarrow c^{\phantom{\dagger}}_\uparrow $& 
  $c^\dagger_\uparrow c^\dagger_\downarrow$ \\ 
%
  $(1-n^{\phantom{\dagger}}_\downarrow)c^{\phantom{\dagger}}_\uparrow$ &
  $(1-n^{\phantom{\dagger}}_\uparrow)c^{\phantom{\dagger}}_\downarrow$&
 $ n^{\phantom{\dagger}}_\uparrow
  c^{\phantom{\dagger}}_\downarrow$& $ n^{\phantom{\dagger}}_\downarrow 
c^{\phantom{\dagger}}_\uparrow$ \\ 
  \rule[-2mm]{0cm}{5mm}
  $(1-n^{\phantom{\dagger}}_\downarrow)c^\dagger_\uparrow$&
  $(1-n^{\phantom{\dagger}}_\uparrow) c^\dagger_\downarrow$& $ 
n^{\phantom{\dagger}}_\uparrow
  c^\dagger_\downarrow$& $ n^{\phantom{\dagger}}_\downarrow
  c^\dagger_\uparrow$\\ \hline
\end{tabular}
\caption{Local operators. The operators in the first two lines have an
  even number of fermionic operators, those in the last two lines
  an odd number of fermionic operators; the usual spin operators are
  $S^z=\sigma^z/2, S^\pm = \sigma^\pm$.
} 
\label{tab:locop}
\end{table}
The local operators of Tab.~\ref{tab:locop} have to be
labelled with a site-index when they are used to write down the
Hamiltonian. Consequently, also products of  local normal-ordered
operators belonging to different sites are  normal-ordered. 
The expectation values of each separate factor
with respect to the reference ensemble  vanish.
It is the prefactor of a product of local operators as defined in
Tab.~\ref{tab:locop} which we will call the \emph{coefficient} of this term.

\begin{figure}[hbp]
  \begin{center}
    \includegraphics[width=0.6\columnwidth]{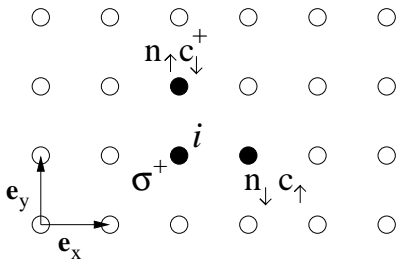}
    \caption{The operator from Eq.~\ref{beispielop} in the 2-dimensional
      square lattice. 
      Filled sites are involved in the hopping process.
      The operators are shown at the sites on which they act.}
    \label{fig:beispielop}
  \end{center}
\end{figure}
On the basis of the normal-ordering and of the
standardized notation introduced above we 
define a measure for the locality of a term, namely its extension $d$.
Consider the following exemplary normal-ordered term  
\begin{eqnarray}
\label{beispielop}
\sum_{i} n^{\phantom{\dagger}}_{i+{\bf e}_y,\uparrow} 
c^\dagger_{i+{\bf e}_y,\downarrow} 
c^{\phantom{\dagger}}_{i+{\bf e}_x,\uparrow} 
n^{\phantom{\dagger}}_{i+{\bf e}_x,\downarrow} 
\cdot {\bf \sigma}_i^+.
\end{eqnarray}
This term represents the hopping of an electron from site 
$i+{\bf e}_x$ to site $i+{\bf e}_y$. The spin of the electron is
flipped in this process, which is compensated by the spin-flip on
site $i$. This term is illustrated in Fig.~\ref{fig:beispielop}. It
contains operators that act on three different sites. They are shown
as filled circles in Fig.~\ref{fig:beispielop}. These sites will be
called the {\it cluster} belonging to this term. The maximal taxi cab 
distance $d$ of two sites in this cluster is the {\it extension} of this term. 
The extension of the term (\ref{beispielop}) in
Fig.~\ref{fig:beispielop} is $d=2$. The
extension of the cluster belonging to a term will be our measure of
its locality. For a truncation scheme, we define a certain set of clusters, 
i.e.\ geometric arrangements of sites which are translated over the whole 
lattice. Those operators  generated in the flow equation
(\ref{fleq}) are kept which can be defined completely on one of the
clusters. If they affect more sites or sites at a larger distance
the operators are discarded.

We emphasize that the use of a set of clusters to define the truncation 
does not turn our approach into a finite cluster calculation. Rather it is 
a self-similar, renormalizing calculation in the thermodynamic limit
where the decision which terms are kept is based on their locality. 
By translational invariance, the operators appear in the whole lattice as 
is denoted by the $\sum_{i}$ in Eq.\ (\ref{beispielop}). 
For large enough clusters, the perturbative results are reproduced
by the self-similar calculation. For simplicity, the differential equation
is set up for the coefficient of a single representative of a certain kind
of term taking advantage of the symmetries of the problem, namely
translational invariance, the point-group symmetries, spin rotation symmetry,
and hermiticity.

The complete prescription for the {\em self similar} CUT calculation reads:
(i) Define a certain set of clusters. Normal-ordered operators on these 
clusters are kept during the calculation. Additional conditions
may be imposed. With this truncation rule the flow equation
(\ref{fleq}) closes. (ii) Calculate $[\eta(\ell),H(\ell)]$ to obtain
the  flow equation (\ref{fleq}) for the coefficients of the terms in
${H}$. Terms not present in the original Hamiltonian start at $\ell=0$ with
initial coefficients zero. (iii) The numerical integration of the flow equation
yields the effective model for $\ell\to\infty$; it
conserves the number of DOs. This step requires that the flow equation
converges for $\ell\to\infty$. If it converges the effective model is
constructed successfully. Generally, non-convergence can be due to
an insufficient approximation or due to the breakdown of the mapping
of the Hubbard model to the $t$-$J$ model.

Our procedure for a self similar CUT calculation will be 
exemplified in the following subsection.
\subsection{Example: The minimal and the NN truncation}
Here we work out a simple example and compare its
results to perturbation theory. 
At first we build a minimal model for the NN coupling
$J$. Starting with $H_t$ and $H_U$ the
generator $\eta$ reads 
\begin{equation}
  \eta(\ell) = [ \hat D , {H}(\ell) ] = [ \hat D, T_{+2}+T_{-2}]
             =2 T_{+2}-2 T_{-2}
\end{equation}
To set up the flow equation we have to calculate
\begin{equation}
  \frac{d}{d\ell}{H}(\ell)=[\eta(\ell),{H}(\ell)]
  =[\eta(\ell),{H}_U(\ell)+H_t(\ell)].
\end{equation}
The first contribution to the flow equation
\begin{equation}
  [ \eta(\ell) , {H}_U(\ell) ] = -2 U(\ell)  T_{+2}-2 U(\ell) T_{-2}
\end{equation}
produces the suppression of terms changing the number of DOs as intended. 
The second contribution reads
\begin{subequations}
\label{mingesamt}
\begin{eqnarray}
[\eta(\ell),{H}_t(\ell)]&=&[2 T_{+2}-2T_{-2},T_0+ T_{+2} + T_{-2}]\nonumber \\
&=&32 t_{+2} t_{-2}\cdot \frac{1}{2} \hat D \label{minhu} \\ 
&& + 16 t_{+2} t_{-2} \sum_{\langle i,j\rangle} {\bf S}_i \cdot {\bf S}_j
\label{minss} \\ 
&& - 4 t_{+2} t_{-2} \sum_{\langle i,j\rangle} 
\bar{n}_i\bar{n}_j \label{minv} \\
&& + 8 t_{+2} t_{-2} \sum_{\langle i,j\rangle} c^\dagger_{i,\uparrow} 
c^\dagger_{i,\downarrow} c^{\phantom{\dagger}}_{j,\downarrow}
c^{\phantom{\dagger}}_{j,\uparrow}
\label{minpair} \\
  && + ... \qquad \textrm{(three site terms)}.\nonumber
\end{eqnarray}
\end{subequations}
In Eq.\ (\ref{mingesamt}) the terms containing operators on three
different sites are omitted.  The term (\ref{minhu}) will renormalize the 
strength of the Hubbard interaction $U$. The next line (\ref{minss}) generates 
the Heisenberg  exchange 
\begin{equation}
  H_{\rm NN}= J_1(\ell) \sum_{\langle i,j\rangle} {\bf S}_i \cdot {\bf S}_j
\end{equation}
with the initial condition $J_1(0)=0$. To get a minimal model for the 
NN exchange  we neglect the terms in (\ref{minv}) and (\ref{minpair}).
The exchange term $H_{\rm NN}$ will not produce a contribution to
$\eta$ since it does not change the number of DOs. As a further
simplification we neglect the terms that would arise from the
commutator $[ \eta, H_{\rm NN}]$. Exploiting that operators related by
hermitian conjugation have the same coefficient we know $t_{+2}=t_{-2}$. Thus 
the differential equations for the minimal truncation read
\begin{subequations}
\begin{eqnarray}
  \label{dglminimal}
  d_\ell U(\ell)&=& 32 t_{+2}(\ell)^2          \\ 
  d_\ell t_{0}(\ell)&=& 0                      \\
  d_\ell t_{+2}(\ell)&=& -2 U(\ell)t_{+2}(\ell)\\ 
  d_\ell J_1(\ell)&=& 16 t_{+2}(\ell)^2 \ ,      
\end{eqnarray}
\end{subequations}
where we use $d_\ell$ as shorthand for $\frac{d}{d\ell}$.
Using the conserved quantity $s=\sqrt{64 t_{+2}(\ell)^2+4 U(\ell)^2}$ we
obtain  for the minimal truncation
\begin{subequations}
\begin{eqnarray}
  \label{solminimal}  
  U(\ell)&=& \frac{s}{2} \tanh(s \ell +C) \\ 
  t_0(\ell)&=& t_0(0)                     \\
  t_{+2}(\ell)&=&t_{-2}(\ell)=(s/8)\sqrt{1-\tanh^2(s\ell+C)} \\
  J_1(\ell)&=& s \left[\tanh(s \ell + C)-2 U_0/s\right]/4    
\end{eqnarray}
\end{subequations}
where $C=\textrm{arctanh} (2U_0/s) $.  
The effective model is obtained in the limit $\ell\to\infty$. 
Since $t_{+2}(\infty)=0$ it contains only \#DO-conserving terms 
\begin{equation}
  \label{hmineff}
  {H}_{\rm eff}= T_0 + U(\infty)\frac{1}{2} \hat D + J_1(\infty)
  \sum_{\langle i,j\rangle} {\bf S}_i \cdot {\bf S}_j.
\end{equation}
Strictly at half-filling, Eq.\ (\ref{hmineff}) reduces to a Heisenberg model 
with effective coupling $J_{1,{\rm eff}}=J_{1}(\infty)$
\begin{subequations}
  \label{J1minimal}
  \begin{eqnarray}
    J_{1,{\rm eff}} &=& \frac{U}{2}
      \sqrt{1+16 \left(t/U\right)^2} - \frac{U}{2}\\
    &=& \frac{4t^2}{U}+\mathcal{O}(t^4/U^3).
  \end{eqnarray}
\end{subequations}
For small $t/U$ this reproduces the well-known second
order result\cite{harri67} $J_1^{(2)} = \frac{4 t^2}{U}$. 
Note that the final results are given in $t$ and $U$ denoting the initial, 
bare values whereas, for clarity,
we use $t_0=t$ and $U_0=U$ to denote  the initial, bare 
values while solving the flow equation.

\begin{figure}[htb]
  \begin{center}
    \includegraphics[width=0.7\columnwidth]{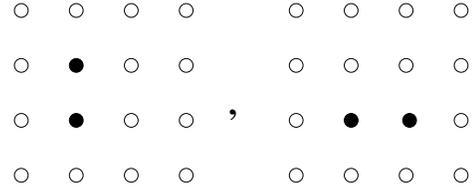}
    \caption{Filled circles indicate the two clusters defining the
      truncation rule of the NN truncation. Only terms fitting on
      these clusters are kept.}  
    \label{fig:NNclusters}
  \end{center}
\end{figure}
The result (\ref{J1minimal}) coincides with the result one gets from 
diagonalization of a two-site cluster for the splitting between
the singlet and the triplet state and thus for $J$.  This is purely
coincidental since taking into account {\it all} operators on NN sites in
the CUT leads to a different result.
The truncation rule for this NN truncation is defined by the two
clusters shown in Fig.~\ref{fig:NNclusters}. All terms are kept that
contain only local operators on two neighboring sites. 
For the effective coupling we find, see Appendix \ref{app:NN},
\begin{eqnarray}
  J_{1,\text{eff}}^\text{NN}= \frac{2}{7} U {\sqrt{1+28 \left(
       t/U\right)^2}} -\frac{2}{7} U\ .
\end{eqnarray}
As expected the result is different from the result
for two sites since the exact diagonalization deals with
a finite system only whereas the  CUT is a
renormalization procedure on the thermodynamic lattice where the
terms kept are chosen according to their locality.

\begin{figure}[htb]
  \begin{center}
    \includegraphics[width=\columnwidth]{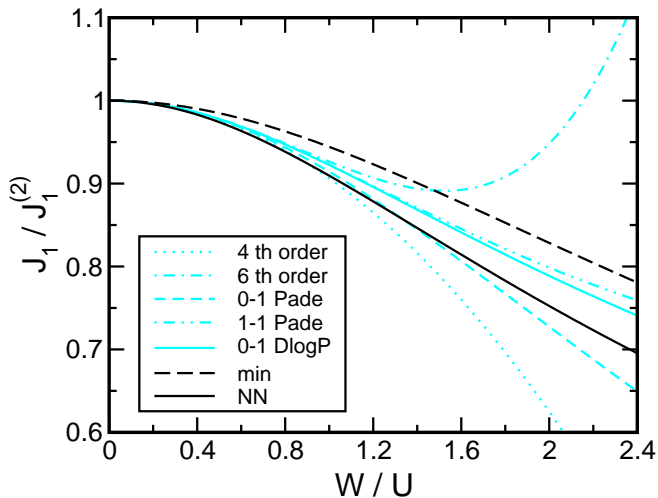}
    \caption{Results for the Heisenberg coupling $J_1$ from the minimal
      and the NN truncation compared to perturbation theory.
      All curves are plotted relative to $J_1^{(2)}=4 t^2/U$.
      The numbers in the legend in front of Pad\'e or DlogP
      indicate the degree of the polynomials in $x=t^2/U^2$ in the
      numerator and denominator of the Pad\'e approximant. Recall $W=8t$.}
    \label{fig:flglundpert}
  \end{center}
\end{figure}
To compare our results to perturbation theory,
we calculated also the 4th and 6th order in $t/U$ for
$J_1$ in perturbation theory \cite{takah77,notiz2}. 
Fig.~\ref{fig:flglundpert} compares the value of the Heisenberg
coupling $J_1$ from the minimal and from the NN truncation to the results
from perturbation theory.  All results are given relative to 
the second order result $J_1^{(2)}=4 t^2/U$. The perturbative results
show strong divergencies even if one includes the 4th and the 6th order. 
The dashed and solid black lines show the solution for $J_1$ from the
minimal and NN truncation, respectively. They give meaningful finite values
for $J_1$ without divergent behavior up to high values of
$W/U$. Already the minimal model correctly reproduces the leading
perturbative behavior $\propto 4 t^2/U$ for small $W/U$.

In addition, Fig.~\ref{fig:flglundpert} shows Pad\'e-approximants 
derived from the series of the perturbation theory. The continuous grey 
line depicts a Dlog-Pad\'e-approximant. The Pad\'e-approximants support that 
the CUT results for $J_1$ lie in the reasonable range of values. But 
neither the CUT nor the Pad\'e results determine $J_1$ reliably for 
values of $W/U\gtrsim 0.8$. Therefore, we 
improve the CUT calculation by taking into account more terms on more complex 
site configurations leading to higher truncation schemes.
The agreement of calculations with various numbers
of operators will be a probe for the accuracy of the results.

\subsection{Higher truncation schemes} \label{sec:hightrunc}
\begin{figure}[htb]
  \begin{center}
    \includegraphics[width=\columnwidth]{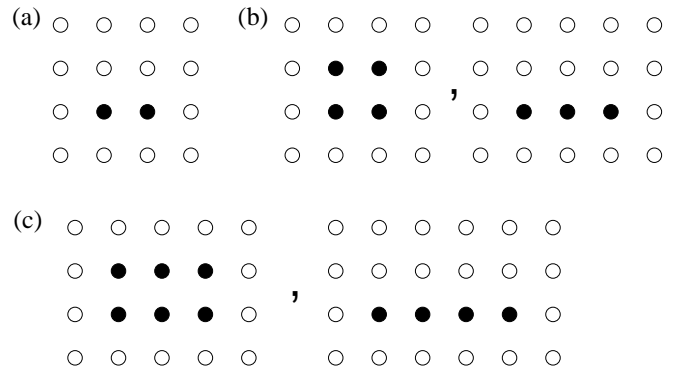}
    \caption{Various clusters defining the operators to be kept in the
      CUT. Only one of various clusters that are related by the point
      group symmetry of the two-dimensional square lattice is shown.}
    \label{fig:cluster}
  \end{center}
\end{figure}

Fig.~\ref{fig:cluster} shows the clusters defining the various
truncation schemes. The sites that belong to the clusters are depicted 
as filled circles within the two-dimensional square lattice. 
Only one representative is shown for different clusters
that are related by the point-group symmetry of the square lattice. 
We use the clusters that are necessary for
a perturbative derivation of the spin Hamiltonian from the Hubbard
model as a guideline for the choice of the clusters defining the
truncation. In second order the NN Heisenberg exchange is the only
effective spin-coupling. For its perturbative derivation a cluster of
two neighboring sites is sufficient. Thus the NN truncation is
defined by the cluster given in Fig.~\ref{fig:cluster}a. 
The cluster in Fig.~\ref{fig:cluster}a represents the two possible
clusters of NN sites on the two-dimensional square lattice. 
In 4th order in the electron hopping $t$ also further hopping
processes contribute to the effective spin-couplings. 
The cluster of four sites on a plaquette of the square lattice and the
linear clusters of three sites cover all relevant hopping processes in
4th order. They are depicted in Fig.~\ref{fig:cluster}b. 
 The double plaquette and the linear chain of four sites depicted
in Fig.~\ref{fig:cluster}c are representatives for the clusters
necessary for accounting fully of the 6th order. 

The calculation for the clusters in
Fig.~\ref{fig:cluster}c was the largest possible. To have an
additional comparison, we also performed
calculations  keeping only the operators  which affect at most
four different sites on the clusters of  Fig.~\ref{fig:cluster}c. 
In summary, the various truncation schemes are 
\begin{itemize}
\item[(i)] {\bf minimal} model: only $H_U$, $H_t$, $H_{\rm NN}$
\item[(ii)] {\bf NN}: cluster used in 2nd order perturbation
  theory, Fig.~\ref{fig:cluster}a.
\item[(iii)] {\bf plaquette} and three-site chain: cluster used in
  4th order, Fig.~\ref{fig:cluster}b. 
\item[(iv)] operators on up to four different sites
  on the clusters of Fig.~\ref{fig:cluster}c. 
\item[(v)] {\bf double-plaquette} and four site chain: clusters used
  in 6th order, see Fig.~\ref{fig:cluster}c. 
\end{itemize}
The NN truncation (ii) yields the leading order result for the NN spin coupling
$J_1$, the plaquette truncation (iii) yields the correct result in 4th order, 
and the double plaquette truncation (v) yields the correct result in 6th order.
 We confirmed that our implementation complies with these checks. 

\begin{table}[htb]
  \begin{center}
    \leavevmode
    \begin{tabular}{|l|c|c|}
      \hline 
      Calculation         &\#operators in ${H}$& \# of terms in DE \\ \hline 
      (i)   minimal       &       4         &        3    \\ 
      (ii)  NN            &       6         &        8    \\ 
      (iii) plaquette     &     172         &    10,364    \\ 
      (iv)                &    2,217         &  1,341,736    \\ 
      (v) double-plaquette&   26,251         & 304,514,721    \\ \hline
    \end{tabular}
    \caption{
      Number of operators in the Hamiltonian reduced by the symmetries given 
      in the main text
      for various approximation schemes. This number 
      coincides with the number of differential equations.
      In the last column the number of bilinear terms on the right hand 
      side the differential equations is listed.}
    \label{tab:calcinfo}
  \end{center}
\end{table}

The computation of the effective model splits into two parts. 
First, one has to set up the flow equation, i.e., one has to calculate the
commutator $[\eta,{H}]$. According to the truncation scheme newly
generated terms are discarded or included in ${H}$, in
$\eta$, and in the calculation of $[\eta,H]$. This is done in a dynamic
fashion in a C${++}$ program. The list of terms considered grows 
in the course of the automatized computation of the commutators. Since we 
consider operators on a finite number of sites there is a
point, when no new terms are generated. At this point the set of
differential equations, the flow equation, is closed.

Second, one has to solve this flow equation. The
differential equations have to be set up only once for each truncation
scheme. Then they must be solved numerically for each set of initial
conditions given by the initial value of $t/U$. 
Since all coefficients
are entangled in the differential equations the numerical integration 
provides all the coefficients -- magnetic exchange couplings, hopping and 
interactions of charge -- of the effective model together.  

To keep the number of terms as small as possible, we made use of the 
point-group symmetries of the
square lattice, i.e., rotations about multiples of $\pi/2$ and
reflections about the axes and diagonals. Additionally, terms which are
related by hermitian conjugation, by particle-hole transformation, and by
flipping $\sigma \to -\sigma$ all spin indices carry the same coefficient. 
Table~\ref{tab:calcinfo} shows the number of terms, reduced by the 
symmetries, in the Hamiltonian of 
each calculation. As $\eta$ and ${H}$ are both linear in the
coefficients of the terms the right hand side of the differential
equations is bilinear in these coefficients. The third column of
Tab.~\ref{tab:calcinfo} lists the number of such bilinear terms in
the differential equation. The number of variables and the number of 
differential equations is growing fast with the size of the cluster
defining the truncation rule. This  renders both the set up of
the flow equation and its numerical integration computationally costly. 

We implemented the CUT calculation on the computer with two
C${++}$-programs. The first sets up the differential equation by
symbolic calculation of the commutator of $\eta$ and ${H}$. 
For a large number of terms in the hamiltonian this task is
very time-consuming. But it is possible to
split the calculation into a number of independent parts that can be 
accomplished by separate programs. For the double plaquette, about
65 separate runs were performed, each with a CPU time below 10 hours
and less than 0.5 Gbyte memory on Sun UltraSPARC workstations.

The second program solves the 
flow equation, i.e., the set of differential equations. 
As the set of equations is large this second 
program is very demanding both in time and in memory space.
Typically 120 hours of CPU time and about 2.5 Gbyte are needed
 on Sun UltraSPARC workstations.

\section{Results}\label{sec:res}
Here we present the results for the five different truncation schemes 
 (i) to (v) described above in  Sect.\ \ref{sec:hightrunc}.
First, we discuss some properties of the 
continuous unitary transformation which enable us to analyze the limitations
of the construction of the $t$-$J$ model. Second, data on the effective
magnetic couplings will be given. Third, data on the effective
charge couplings will be shown.

\subsection{Properties of the CUT}
\begin{figure}[htb]
  \begin{center}
    \includegraphics[width=\columnwidth]{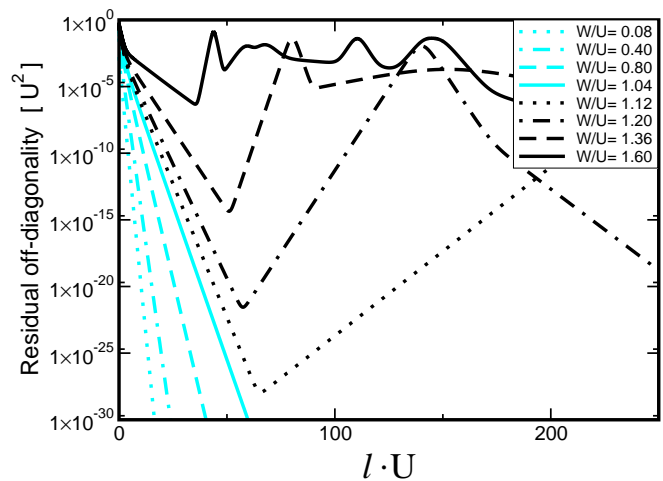}
    \caption{Evolution of the residual off-diagonality in the
      double-plaquette calculation (v) for various values of $W/U$. }
    \label{fig:offdiagonal}
  \end{center}
\end{figure}
The CUT is induced by operators that change the number of DOs. They are 
off-diagonal in the number of DOs. We define the {\it residual off-diagonality}
as the sum per site of the squares of all coefficients that belong to operators
that change the number of DOs. In the calculation the Hamiltonian is written 
in terms of products of the local operators in Table \ref{tab:locop}. This 
implies the  definition of the coefficient as the prefactor of such a product.
  
The residual off-diagonality measures the extend to which the unitary 
transformation has eliminated the terms that  change the number of DOs. 
In a numerical integration of the differential
equation the residual off-diagonality will not vanish exactly. 
Therefore, we  define a small but non-zero value and stop the
flow when the residual off-diagonality falls below this value. 
Fig.\ \ref{fig:offdiagonal} displays the evolution of the
residual off-diagonality in the double-plaquette calculation (v)
during the flow, parametrized by the variable $\ell U$, 
see Sect.~\ref{sec:cut}, for various values of $W/U$. 
At $\ell=0$ the operators in $T_{+2}$ and $T_{-2}$ in (\ref{eq:ham}) 
contribute to the residual off-diagonality.
Written with the operators in Table \ref{tab:locop} these are 
$4*2=8$ terms (hopping forth and back in x- and y-direction, with spin
up or down) in $T_{+2}$ and $T_{-2}$ each. Therefore the curves in 
Fig.\ \ref{fig:offdiagonal} for the residual off-diagonality start at
$\ell=0$ with the value $8 t_{+2}^2+8 t_{-2}^2=16 t^2$. 

For small values of $W/U$ the residual off-diagonality shows exponential
decay.  For larger values
of $W/U$ the convergence is slower.  For $W/U=1.12$ it shows for the 
first time pronounced non-monotonic behavior. The dotted black curve,
corresponding to $W/U = 1.12$, falls below $10^{-27}$ to rise again up to
$10^{-3}$ for $\ell\cdot U\approx 270$. 
Non-monotonic behavior occurs for all values $W/U\ge 1.12$; it
 can become a problem since we have to compare the residual
off-diagonality to a small but finite value to decide at which value
of $\ell$ to stop the integration of the differential equations.
But the values of the dominant coefficients of the effective model
are reached already at small $\ell$. They are not affected
noticeably by different cutoffs for the residual off-diagonality.

Most importantly, one has to check whether the mapping to the effective
model is possible at all. Below we will 
calculate the apparent charge gap $\Delta_\text{app}$
within the effective model. This is the minimal energy of a DO, e.g.\ a
hole, propagating through a paramagnetic spin background after
the CUT. Since the paramagnetic spin background is not the
ground state the apparent charge gap $\Delta_\text{app}$
is \emph{not} the true charge gap. The apparent charge gap
measures the separation of energy scales. This separation
governs the convergence of the mapping as we  perform it. 
We will discuss this issue and its physical significance in more detail 
in the Discussion.

We find that the apparent charge gap $\Delta_\text{app}$
becomes  negative (!) for values of $W/U \gtrsim 0.92$.
Once excitation energies become negative the CUT does not work
because the limit $\ell \to \infty$ does not exist any longer.
The \#DO-changing processes can no longer be eliminated because
they constitute no longer vacuum fluctuations around a stable vacuum.
The negativity of excitation energies signals that the chosen reference
ceases to be a physically reasonable reference.

The non-monotonic behavior of the residual off-diagonality 
appears only in a parameter range where the mapping is no longer possible. 
For parameters $W/U\lesssim 0.9$, where the mapping is possible according
to $\Delta_\text{app} >0$, we
find an exponential decrease of the residual off-diagonality down to 
values where the calculation is limited by the numerical accuracy of
the computer implementation. 
A pronounced increase of the residual off-diagonality was only 
found in the double-plaquette calculation (v). The calculations
(i)--(iv)  show an exponential drop of the residual
off-diagonality for all values $W/U$.  

\begin{figure}[htb]
  \begin{center}
    \includegraphics[width=\columnwidth]{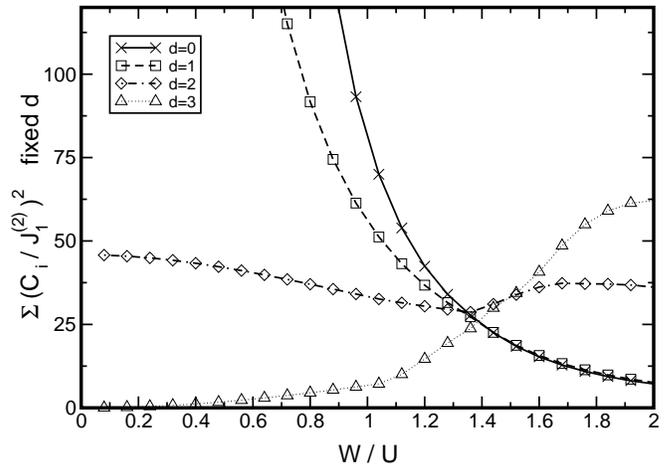}
    \caption{Distribution of coefficients as function of $W/U$ according to 
      the spatial extension $d$ of the respective terms in the double-plaquette
      calculation (v). The sum of the squares of the coefficients are
      plotted relative to the second order Heisenberg exchange
      $J_1^{(2)}=4t^2/U$.} 
    \label{fig:ainachgroesse}
  \end{center}
\end{figure}
Next, we analyze the distribution of coefficients over operators 
of various extensions $d$, see Sect.\ \ref{sec:cut},
and the effect of truncating the proliferating terms.  
Fig.~\ref{fig:ainachgroesse} shows the sum of the squares of all 
coefficients $C_i$ of
terms with the definite extension $d$ at the end of the flow.
They are given  as functions of $W/U$. The results are 
given in units of $J_1^{(2)}= 4 t^2/U$ which
is a natural scale for the parameters defining the
effective model. The only operator with extension zero is the Hubbard
repulsion term ${H}_U$. For increasing $W/U$ the relevance
of the most local terms with $d=0$ and $d=1$ decreases, whereas terms
with larger extension become more and more important.  Note especially the
increase of the coefficients of the operators with maximal 
extension $d=3$. 

The distribution of coefficients indicates that up to $W/U\approx 1.2$
all important operators are included in our calculation. Beyond
$W/U \gtrsim 1.2$ the coefficients with extension $d=3$ become more
important. Thus terms with larger extension
would have sizeable coefficients and should be taken into account. 
This indicates that the approximations become definitely insufficient beyond
$W/U \gtrsim 1.2$.
\begin{figure}[htb]
  \begin{center}
    \includegraphics[width=\columnwidth]{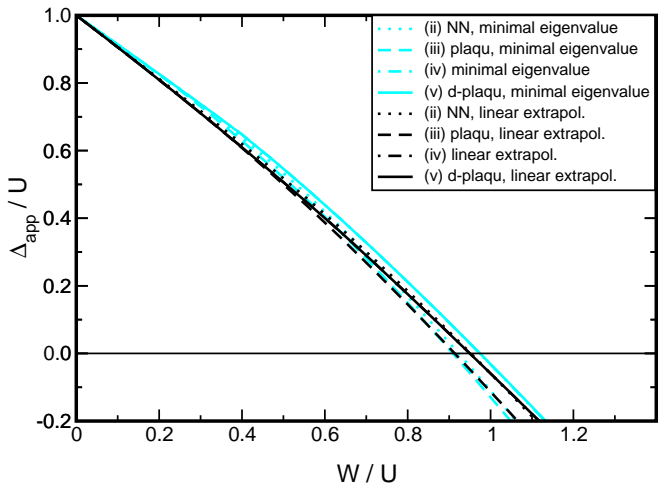}
    \caption{Apparent charge gap as calculated with Lanczos
      diagonalization. Grey lines show the minimal eigen values, black
      lines the result of a linear extrapolation.}
    \label{fig:gap}
  \end{center}
\end{figure}

To obtain clearer information on the
applicability of the mapping and on the accuracy of the
approximations we examine the apparent charge gap $\Delta_\text{app}$.
To this end, the dispersion $E_k$ of a single DO moving in the 
paramagnetic spin background is determined by a Lanczos calculation 
\cite{petti85,viswa94}. In this calculation we treat the operators
as vectors as in the projection technique
\cite{fulde93}. The effective Hamiltonian acts by commutation 
as a superoperator on the operators. This is conveniently denoted by the 
Liouville operator $\mathcal{L} := [H, \cdot]$. The  ``scalar product''
$(A|B)$  is defined by $\text{Tr}(A^\dagger B \hat \rho_0)$ relative
to the paramagnetic reference ensemble (\ref{refensemble0}).
Note that this scalar product is only positive semi-definite since
there can be operators whose norm vanishes.

The projection formalism is the appropriate concept for this calculation
since we are not dealing with simple states in a Hilbert space but with
ensembles which are derived from the paramagnetic one (\ref{refensemble0})
by the application of operators.  These operators induce deviations from the
reference ensemble which constitute excitations. As usual in the Heisenberg
picture, the dynamics of operators in the frequency domain
is captured by the commutation with the Hamiltonian.

We start the Lanczos approach from the operator ${\mathcal O}_0$ which 
generates a single DO with momentum $k$ 
\begin{equation}
\label{initoper}
  {\mathcal O}_0 := \frac{1}{\sqrt{N}} \sum_{\bf{r}} e^{i \bf{k}\bf{r}}
n^{\phantom{\dagger}}_{\bf{r},\downarrow}c^\dagger_{\bf{r},\uparrow}
\end{equation}
where the number of sites is denoted by $N$ and the sum
extends over the whole lattice. A set of orthogonal operators
is generated by the iterative application of $\mathcal{L}$
\begin{equation}
  {\mathcal O}_{n+1}=
  {\mathcal L}{\mathcal O}_{n}  - a_n{\mathcal O}_{n} -b^2_n
{\mathcal O}_{n-1}\ ,
\end{equation}
where
\begin{subequations}
\begin{eqnarray}
  a_n&=& ({\mathcal O}_{n}|{\mathcal L}{\mathcal O}_{n}) / 
  ({\mathcal O}_{n}|{\mathcal O}_{n}) \ ,\\
  b_{n+1}^2&=&({\mathcal O}_{n+1}|{\mathcal O}_{n+1}) / 
  ({\mathcal O}_{n}|{\mathcal O}_{n}) ,\quad b_0=0\ .
\end{eqnarray}
\end{subequations}
Note that the application of $\mathcal{L}$ derived from the effective
Hamiltonian \emph{after} the CUT
 does not change the number of DOs. Hence its action on the initial 
vector is (i) to shift the DO and/or (ii) to change the spin background.
The iteratively built basis  $\{ {\mathcal O}_{n} \}$ describes
a single DO (charge excitation) at momentum $\bf{k}$ including its 
magnetic dressing. In this basis  the Liouville operator
is a tridiagonal matrix where the $a_n$ are the diagonal matrix elements 
and the $b_n$  are the elements on the secondary diagonal.

The lowest energy in this subspace defines the 
(apparent) dispersion $E_k$ of a single charge excitation relative to
the paramagnetic ensemble. Hence the evaluation of the
lowest accessible energy in any truncated subspace of $\{ {\mathcal O}_{n} \}$
provides an upper bound to  $E_k$.
The apparent charge gap $\Delta_\text{app}$ is finally given by
\begin{equation}
  \Delta_\text{app}= 2 \min_k {E_k}\ ,
\end{equation}
where the factor of 2 is put to account for the creation of
a particle and a hole. For vanishing hopping $\Delta_\text{app}=U$ holds.
 The apparent charge gap $\Delta_\text{app}$ 
is not the true charge gap because the paramagnetic spin background
is not the true ground state. But $\Delta_\text{app}$ is a measure of the 
separation of energies between sectors of different double occupancy.

Numerically, one has to restrict the above procedure to truncated basis
sets. In practice, we construct a sequence of basis sets, labelled
by the integer $n$, by applying the Liouville operator
$\mathcal{L}_0 := [T_0, \cdot]$ (cf.\ Eq.~\ref{eq:T0}) iteratively to the 
initial operator (\ref{initoper}).
The first basis $n=1$ consists only of the single operator
${\mathcal O}_0$. The subsequent basis sets are generated by applying  $T_0$
$(n-1)$ times. All the components which are products of the local
operators in Tab.\ \ref{tab:locop} are treated as independent vectors of the 
basis. For each basis set the lowest eigen value of the
matrix representation of the Hamiltonian in this basis yields
an upper bound to the lowest energy $E_k$ of a single DO.

The calculation was done for $\bf{k}$-values on the high symmetry lines in
$\bf{k}$-space from $(0,0)-(\pi/a,\pi/a)-(\pi/a,0)-(0,0)$. 
Figure \ref{fig:gap} shows the results of the Lanczos calculation. 
The computational effort grows with the number of terms 
in the Hamiltonian. For the effective Hamiltonians (iv) and (v)
it was only possible to do the 
calculation up to the 
basis set $n=6$, whereas for the NN (ii) and 
plaquette (iii) Hamiltonians the $n=7$ calculation was feasible.
In Fig.~\ref{fig:gap} grey lines show the smallest eigen value 
found for all momenta $k$ in the largest possible Lanczos calculation.
An additional estimate for the gap is found by extrapolating
the lowest eigen values at momentum $(\pi/a,\pi/a)$ for each basis set
as function of  $1/n$ to $n\to \infty$. 

The lowest eigen value at momentum $(\pi/a,\pi/a)$ is usually also 
the minimum of the dispersion except for $n=5,6$ or $7$ where the
whole dispersion becomes very flat. The result from 
a fit linear in $1/n$ is included in Fig.~\ref{fig:gap} by black lines. 
For the minimal eigen values as well as for the extrapolated ones
the result for the Hamiltonian (iv) is covered  by 
the double-plaquette result (v), since the coefficients of the 
Hamiltonians nearly coincide. 
The lowest eigen values  for the Hamiltonians  (iv) and (v) are larger than
the lowest eigen values for the  Hamiltonians (ii) and (iii)
because larger values of $n$ were accessible for the latter.

A first guess for the gap is  $\Delta_{\rm app} /U = 1 - W/U$, 
see e.g.\ Ref.\ \onlinecite{gebha97}. So one  expects that the gap closes 
around $W/U = 1$. The apparent charge gap calculated within the effective 
model using Lanczos-diagonalization displays a similar behavior. 
The gap decreases almost linearly from $\Delta_{\rm app} / U =1$ at
$W/U=0$.  There is, however, a certain downward curvature so that it closes
at about  $W/U\approx 0.9$.  Taking into account even larger basis sets might 
 further lower the ratio $W/U$ where the gap closes. 
The linear extrapolation corroborates the results from the linear
eigen values to a certain extent. But we could not identify a
clear asymptotic behaviour for large values of $n$. Other extrapolation
schemes, for instance as function of $1/\sqrt{n}$, point towards
an earlier closing of the apparent charge gap.
 Therefore the value $W/U\approx 0.9$ for the closing of the gap must be seen 
as a rough estimate. The determination of the  power law by which the
gap closes is beyond the scope of the present paper.

The above findings are discussed in more detail in Sect.\ \ref{sec:dis}. 
In the sequel, we have to keep in mind that the mapping to the effective
model conserving the number of DOs is not possible for $W/U\gtrsim 0.9$.
For completeness, we will present the coefficients of 
the generalized $t$-$J$ model up to values  $W/U=1.6$ to illustrate in which
way the results are affected by the breakdown of the mapping. 
As a reminder of the breakdown,  the definitely 
unphysical region beyond $W/U\approx 1$ will be shaded in the figures.

\subsection{Effective spin model}\label{sec:resspin}
\begin{figure}[htb]
  \begin{center}
    \includegraphics[width=\columnwidth]{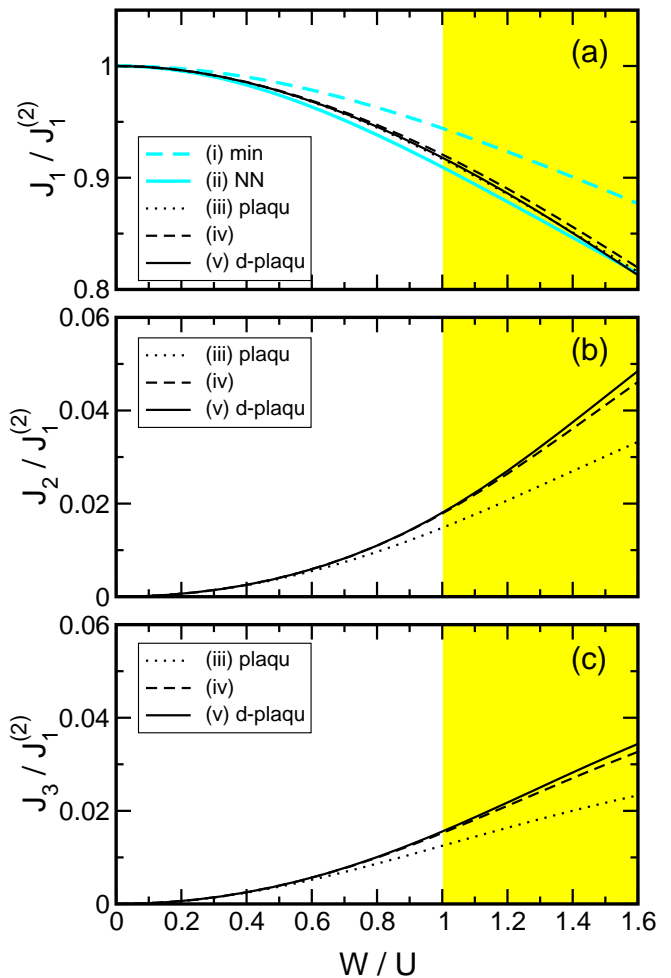}
    \caption{Couplings  of two-spin operators 
      relative to $J_1^{(2)}$ as function of $W/U$.}
    \label{fig:zweispin}
  \end{center}
\end{figure}

Strictly at half-filling,
 the effective model can be cast into the form of a pure
spin Hamiltonian. There are various two-spin couplings of the
Heisenberg-type
\begin{equation}
  H_{\textrm{2-spin}}=  \sum_{i,j}J_{|j-i|} {\bf S}_i \cdot {\bf S}_j.
\end{equation}
This type of exchange may exist for sites $i$ and $j$ of any distance but it
is of course largest for adjacent sites. Fig.~\ref{fig:zweispin} displays the
coefficients of various two-spin couplings. The Heisenberg
NN exchange $J_1$ is shown in Fig.~\ref{fig:zweispin}a, the exchange
couplings between second and third nearest neighbors, $J_2$ and $J_3$,
are depicted in Fig.~\ref{fig:zweispin}b and \ref{fig:zweispin}c,
respectively. The coefficients are 
shown relative to $J_1^{(2)}=4t^2/U$. The coupling $J_1$ is  only
renormalized slightly to lower values $\approx 0.92 J_1^{(2)}$ for
$W/U=1$ contrary to what one might have expected from the strong influence
of higher terms in perturbation theory. Already the NN 
calculation reproduces the leading order correctly and
yields a very reasonable result for larger values of $W/U$,
see also Fig.~\ref{fig:flglundpert} and the discussion thereof.
The convergence of the 
various calculations shows that the terms retained are sufficient to
determine $J_1$ to very good accuracy. Note that the dotted curve of the
plaquette calculation (iii) is almost hidden by the double-plaquette
result (v). 
\begin{figure}[htb]
  \begin{center}
    \includegraphics[width=\columnwidth]{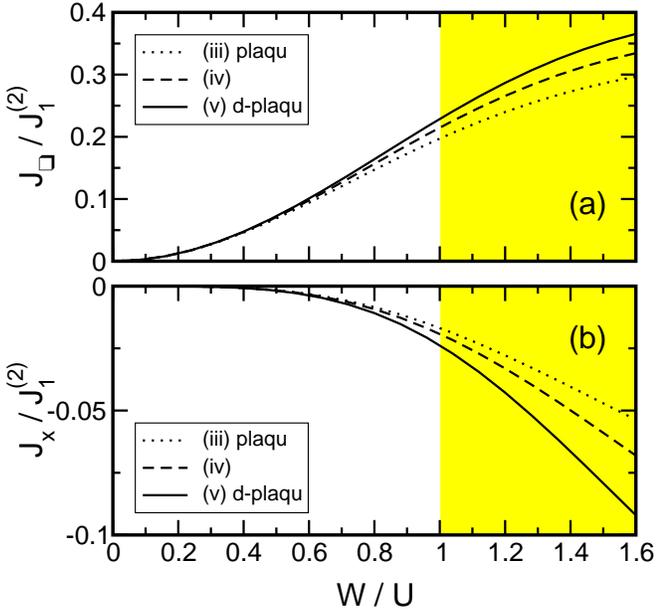}
    \caption{Couplings  of four-spin operators 
      relative to $J_1^{(2)}$ as function of $W/U$.
      }
    \label{fig:vierspin}
  \end{center}
\end{figure}

The couplings $J_2$ and $J_3$ behave like $(t/U)^4$ for
small $W/U$. Their absolute convergence is as good as for $J_1$.
But due to their small values -- both are
smaller than $0.02 J_1^{(2)}$ for $W/U < 1$ --
the relative differences between calculations on different
clusters are still discernible.

There are not only exchange terms between two spins but also between
four or more spins. Exchange between an odd number of spins is not 
generated since such terms would change sign under time reversal. 
The first term that one encounters in perturbation
theory \cite{takah77} in 4th order is the four-spin part of the
ring-exchange which takes the form
\begin{eqnarray}
  {H}_{\Box}&=&J_{\Box}\sum_{\langle i,j,k,l\rangle}
 \Big[({\bf S}_i\cdot{\bf S}_j)({\bf S}_k\cdot{\bf S}_l)+
({\bf S}_i\cdot{\bf S}_l)({\bf S}_j\cdot{\bf S}_k) \nonumber  \\
&&\phantom{J_{\Box}\sum_{\langle i,j,k,l\rangle}} -
({\bf S}_i\cdot{\bf S}_k)({\bf S}_j\cdot{\bf S}_l)\Big]
\end{eqnarray}
where $\langle i,j,k,l\rangle$ label the sites around a plaquette of the 
square lattice in cyclic order. In 6th order an additional independent
four-spin operator reads  
\begin{equation}
  {H}_{\times}=
J_{\times} 
\sum_{\langle i,j,k,l\rangle}({\bf S}_i\cdot{\bf S}_k)({\bf S}_j\cdot
{\bf S}_l). 
\end{equation}
The coefficients $J_\Box$ and $J_\times$ are plotted in
Figs.~\ref{fig:vierspin}a and \ref{fig:vierspin}b. The ring exchange
$J_\Box$ is the most important modification of the pure NN Heisenberg
model. Its value is larger than $0.2 J_1^{(2)}$ for $W/U=1$.
The relevance of the ring-exchange was also confirmed
experimentally in the ${\rm CuO_2}$-planes of ${\rm La_2CuO_4}$
\cite{colde01b} and theoretically in the derivation of the effective
spin model from a three-band model \cite{schmi90,mizun99,mulle02a}.
The influence of ring-exchange on the Raman line shape was computed
in spin wave theory\cite{katan02}.

It is worth noting that the coefficients of the effective spin model  
are smooth functions of $W/U$ up to $W/U=1.6$. 
The vanishing of the apparent gap $\Delta_{\rm app}$ and the slow 
drop of the residual off-diagonality for $W/U\gtrsim 1$ 
do not give rise to an anomalous behavior of the effective spin 
coefficients. The convergence of the results for values
$W/U \lesssim 1$ is very good, i.e., the couplings do not
change much as function of the cluster size for the larger clusters.

\subsection{Effective charge model}
\label{sec:rescharge}
The unitary transformation deals also with the motion and interaction
of charges. 
In the course of the transformation charge excitations (DOs) are
created virtually. After the transformation, DOs can be introduced externally 
by doping or thermal fluctuations. Note that generically the effective model 
displays a finite $U$ term. So DOs are not suppressed.
But due to the disentanglement of the sectors
of different number of DOs there is no direct influence of these sectors
on one another. For instance, the magnetism strictly at half-filling is
not influenced by the physics of the sector with a particle-hole pair 
(two DOs in our counting). 

But thermal expectation values are generically influenced
by the  presence of sectors of different number of DOs. Also spectral
properties are influenced by the presence of these sectors. The spectral
weight will be distributed over a number of sectors. There
are Hubbard bands beyond the lower (one hole) and the upper (one electron) 
Hubbard band, for instance a trans-upper Hubbard band characterized by
two electrons and a hole relative to the paramagnetic reference ensemble.
But preliminary estimates of the spectral weight in this band indicate that 
it is below the order of one percent. Yet it is a strong point of the approach
chosen that the concept allows to discuss such subtle effects.

In the present paper, we focus on the effective Hamiltonian leaving
the determination of spectral weights to future work.
We will present the results for the most important terms concerning
the dynamics of charges. In order not to be lost in a proliferating
number of terms we will present results for terms which are at most
of order $t^2/U$ in the limit of weak hopping.

First, we address the terms which move a single DO, i.e.\ various
hopping processes.
Besides the NN hopping $T_0$, the next relevant 
hopping processes are those between next-nearest neighbor (NNN)
sites and third-nearest neighbor sites (3NN). The NNN sites lie on
diagonally opposite corners of the plaquettes of the two-dimensional
square lattice. All hopping processes and interactions that appear in
second order in $t/U$ will be shown. The NNN hopping $T'_0$ reads
\begin{eqnarray}
\label{eq:T0s}
  T'_0&=& t'\sum_{{\langle \langle i,j\rangle\rangle}; {\sigma}}
  \Big[(1-n^{\phantom{\dagger}}_{i,\sigma}) 
c^\dagger_{i,\overline{\sigma}}c^{\phantom{\dagger}}_{j,\overline{\sigma}} 
(1-n^{\phantom{\dagger}}_{j,\sigma})\nonumber \\ 
  &&\phantom{t_0\sum_{{ \langle i,j\rangle, \sigma}
      }\Big[}
  -n^{\phantom{\dagger}}_{i,\sigma} c^\dagger_{i,\overline{\sigma}}
c^{\phantom{\dagger}}_{j,\overline{\sigma}} n^{\phantom{\dagger}}_{j,\sigma}
+ \text{h.c.}
  \Big]
\end{eqnarray}
where $\langle \langle i,j\rangle\rangle$ stands for NNN sites and 
$\overline{\sigma}=-\sigma$. The spin-dependent NNN hopping $T'_{s,0}$
reads 
\begin{eqnarray}
  T'_{s,0}&=& t'_s \sum_{\langle i,k,j\rangle; \alpha,\beta}
  \biggl\{
  \big[(1-n^{\phantom{\dagger}}_{i,\alpha}) c^\dagger_{i,\overline{\alpha}}
  \boldsymbol{\sigma}_{\overline{\alpha},\overline{\beta}}
  c^{\phantom{\dagger}}_{j,\overline{\beta}} 
  (1-n^{\phantom{\dagger}}_{j,\beta}) \big] \boldsymbol{\cdot} {\bf S}_k
  \nonumber \\ 
  && \hspace*{1cm}
  +\big[ n^{\phantom{\dagger}}_{i,\alpha} c^\dagger_{i,\overline{\alpha}} 
  \boldsymbol{\sigma}_{\overline{\alpha},\overline{\beta}}
  c^{\phantom{\dagger}}_{j,\overline{\beta}}
  n^{\phantom{\dagger}}_{j,\alpha}\big] \boldsymbol{\cdot}
  {\bf S}_k +\text{h.c.} \biggr\}
\end{eqnarray}
where $\langle i,k,j\rangle$ stands for three sites wherein $i$ and $j$ 
are NNN and $k$ is a nearest neighbor to both $i$ and $j$. 
The symbol $\boldsymbol{\sigma}= (\sigma_x,\sigma_y,\sigma_z)$ stands
for the vector made out of the three Pauli-matrices; ${\bf S}_k$
represents the usual $S=1/2$ spin vector at site $k$.
Hopping and spin-dependent hopping between 3NN is also produced in
second order in $t/U$. They read
\begin{eqnarray}
  T''_0&=& t''\sum_{{\langle \langle \langle i,j\rangle\rangle\rangle}; 
{\sigma}}  \Big[(1-n^{\phantom{\dagger}}_{i,\sigma}) 
c^\dagger_{i,\overline{\sigma}}c^{\phantom{\dagger}}_{j,\overline{\sigma}} 
(1-n^{\phantom{\dagger}}_{j,\sigma})\nonumber \\ 
  &&\phantom{t_0\sum_{{ \langle i,j\rangle, \sigma}
      }\Big[}
  -n^{\phantom{\dagger}}_{i,\sigma} c^\dagger_{i,\overline{\sigma}}
c^{\phantom{\dagger}}_{j,\overline{\sigma}} n^{\phantom{\dagger}}_{j,\sigma}
+ \text{h.c.}
  \Big]
\end{eqnarray}
where $\langle \langle \langle i,j\rangle \rangle \rangle$ stands for
3NN sites and
\begin{eqnarray}
  \label{eq:Tpps}
  T''_{s,0}&=& t''_s \sum_{\langle \langle i,k,j\rangle\rangle; \alpha,\beta}
  \biggl\{
  \big[(1-n^{\phantom{\dagger}}_{i,\alpha}) c^\dagger_{i,\overline{\alpha}}
  \boldsymbol{\sigma}_{\overline{\alpha},\overline{\beta}}
  c^{\phantom{\dagger}}_{j,\overline{\beta}} 
  (1-n^{\phantom{\dagger}}_{j,\beta}) \big] \boldsymbol{\cdot} {\bf S}_k
  \nonumber \\ 
  && \hspace*{1cm}
  +\big[ n^{\phantom{\dagger}}_{i,\alpha} c^\dagger_{i,\overline{\alpha}} 
  \boldsymbol{\sigma}_{\overline{\alpha},\overline{\beta}}
  c^{\phantom{\dagger}}_{j,\overline{\beta}}
  n^{\phantom{\dagger}}_{j,\alpha}\big] \boldsymbol{\cdot}
  {\bf S}_k +\text{h.c.}  \biggr\}
\end{eqnarray}
\begin{figure}[htb]
    \includegraphics[width=0.8\columnwidth]{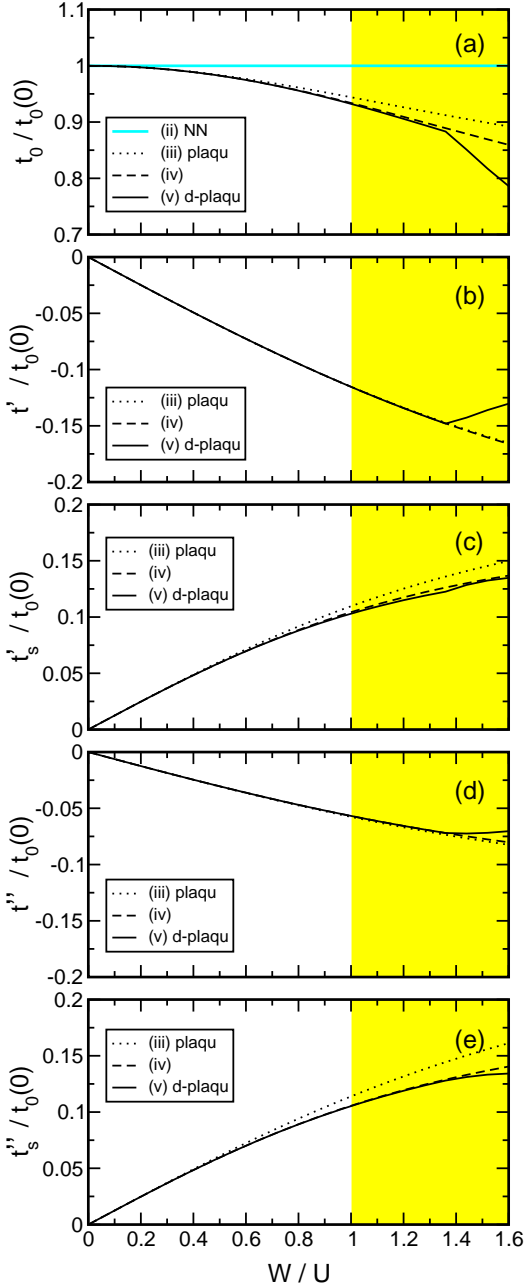}
    \caption{Effective hopping coefficients between (a) NN sites, (b)
      NNN sites, (c) spin-dependent NNN hopping, (d) 3NN
      sites and (e) spin-dependent 3NN hopping. }
    \label{fig:t}
\end{figure}
where $\langle\langle i,k,j\rangle\rangle$ stands for three sites wherein $i$ 
and $j$ are 3NN and $k$ is a nearest neighbor to both $i$ and $j$, so 
$k$ is just the site between $i$ and $j$. 
All these hoppings (\ref{eq:T0s})-(\ref{eq:Tpps}) conserve the number of DOs. 
Figs.\ \ref{fig:t}a-e show the coefficients of
the NN hopping $t_0$ ($T_0$ from Eq.\ (\ref{eq:T0})), NNN hopping $t'$,
spin-dependent NNN hopping $t'_s$, 3NN hopping $t''$ and
spin-dependent 3NN hopping $t''_s$, respectively. They are
shown in units of the unrenormalized value $t_0(0)=t$ for the NN hopping.

The NN hopping $t_0$ shown in Fig.\ \ref{fig:t}a remains unchanged from its
initial value for the NN truncation. 
Including more terms reduces the effective $t_0$.  
At $W/U\approx1.36$, the double plaquette calculation (v) shows a 
sudden change of slope. This is related to the non-monotonic
behavior of the residual off-diagonality. But 
the value $W/U\approx1.36$ is already beyond the point $W/U\approx 0.9$
where the apparent gap $\Delta_\text{app}$
becomes negative and the mapping is no longer possible. 
The signs of $t'$ (NNN hopping) and of $t''$ (3NN hopping) are
opposite to the sign of $t$ (NN hopping). 
The value of $t'$ is approximately two times $t''$ because it is
generated from the two possible hopping processes to go from one corner
of a plaquette to the opposite corner. In contrast,  there is only one
 process to generate  $t''$ in second order. The results for the calculations
(iii)--(v) lie almost on top of each 
other for the coefficients $t$, $t'$ and $t''$ until spurious behavior
sets in at larger values of §W/U§. 

The coefficients of the spin-dependent hoppings $t'_{s}$ and $t''_{s}$ 
are shown in Fig.~\ref{fig:t}c and e. They increase quadratically for
small $W/U$ just like $t'$ which leads to the linear behavior in units
of $t_0$ depicted in Fig.\ \ref{fig:t}. These hoppings are terms which
concern three sites. Their coefficients have approximately the
same value since it does not matter very much whether the
three sites are aligned linearly or form an angle of 90 degrees. 
We emphasize that both coefficients  $t'_{s}$ and $t''_{s}$  are of similar 
size as $t'$.  This implies  that the commonly
used phenomenological description in terms of a $t$-$t'$-$J$ model is
not justified. If an extension of the simplemost $t$-$J$ model is used
the NNN and 3NN hoppings, $t'$ and $t''$, \emph{and} the
spin-dependent hoppings $t'_s$, $t''_s$ should all be included.  

\begin{figure}[htb]
  \begin{center}
    \includegraphics[width=\columnwidth]{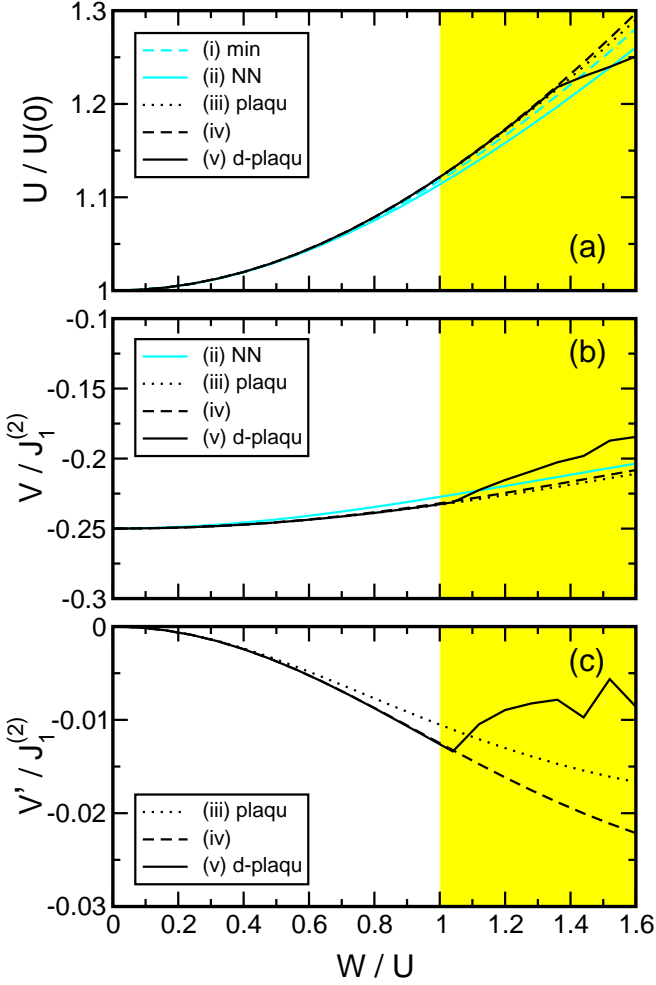}
    \caption{Hubbard repulsion $U$ (a) and NN interaction $V$ (b) and 
      NNN interaction $V'$ (c) between holes or double occupancies.}
    \label{fig:U}
  \end{center}
\end{figure}
\begin{figure}[t]
    \includegraphics[width=0.8\columnwidth]{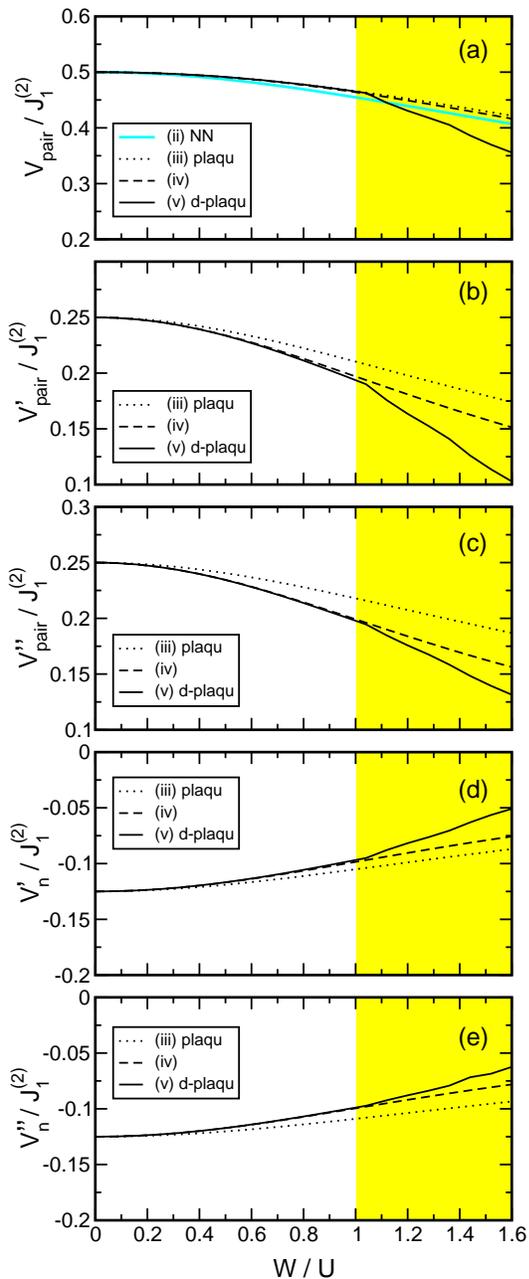}
    \caption{Further coefficients of interactions of two DO that
      are of second order in $t/U$. For explanation, see text. }
    \label{fig:V}
\end{figure}

Figures \ref{fig:U} and \ref{fig:V} show the coefficients for terms
that describe interactions between charges. 
In Fig.~\ref{fig:U}a, the Hubbard repulsion $U$  
is shown in units of its unrenormalized value. All
calculations show a slight increase of the effective $U$ with $W/U$.
It is interesting to note that this basic quantity neither diverges
nor stays constant in the process of the disentanglement of
the various sectors of double occupancy.

All processes discussed in the sequel have the property to
be active only if at least two DOs are present. In our counting
this means that at least two holes or two doubly occupied sites
or one hole and one doubly occupied site must be present. In this
sense, the following processes represent two-particle interactions.
The density-density interactions for NN sites is described by  
\begin{eqnarray}
  H_V &=& V \sum_{\langle i,j\rangle} \bar{n}_i  \bar{n}_j                   
\end{eqnarray}
where $\bar{n}_i=n^{\phantom{\dagger}}_{i,\uparrow} + 
n^{\phantom{\dagger}}_{i,\downarrow}-1$. 
Similarly, the operator for density-density interactions for  NNN sites reads
\begin{eqnarray}
  H_{V'} &=& V' \sum_{\langle \langle i,j\rangle\rangle} \bar{n}_i  \bar{n}_j. 
\end{eqnarray}
The coefficients $V$ and $V'$ are shown in Fig.~\ref{fig:U}b
and \ref{fig:U}c. They are given in units of the second
order scale $J_1^{(2)}=4t^2/U$. The attractive interaction $V$ is proportional 
to $-t^2/U= - J_1^{(2)}/4$ as expected from perturbation theory. It
represents a sizeable contribution at all $W/U$. In contrast, the
$V'$ coupling NNN sites, which is of fourth order in $t/U$, turns out
to be very small \cite{dagot94b}. This fact justifies our choice 
to focus otherwise  on the terms which appear already in second order in $t$.

There are two additional types of operators that are created in second
order in $t/U$. The first type consists of operators which destroy or create
two electrons at the same site. Figure \ref{fig:V}a shows the
coefficient $V_\text{pair}$ of 
\begin{equation}
{H}_\text{pair}= V_\text{pair}
  \sum_{\langle i,j\rangle} c^\dagger_{i,\uparrow} c^\dagger_{i,\downarrow} 
  c^{\phantom{\dagger}}_{j,\downarrow} c^{\phantom{\dagger}}_{j,\uparrow}.
\end{equation}
Figure  \ref{fig:V}b and c show the coefficients $V'_\text{pair}$ and
$V''_\text{pair}$  of 
\begin{eqnarray}
{H}'_\text{pair}&=& V'_\text{pair}
  \sum_{\langle i,k,j\rangle,\sigma} \Big[
  c^\dagger_{k,\sigma} c^\dagger_{k,\bar{\sigma}} 
  c^{\phantom{\dagger}}_{i,\bar{\sigma}}n^{\phantom{\dagger}}_{i,\sigma}
  c^{\phantom{\dagger}}_{j,\sigma}(1-n^{\phantom{\dagger}}_{j,\bar{\sigma}})
  \nonumber \\
  && +c^\dagger_{k,\sigma} c^\dagger_{k,\bar{\sigma}} 
  c^{\phantom{\dagger}}_{i,\bar{\sigma}}(1-n^{\phantom{\dagger}}_{i,\sigma})
  c^{\phantom{\dagger}}_{j,\sigma}n^{\phantom{\dagger}}_{j,\bar{\sigma}}
  +\text{h.c.}\Big],\\
  {H}''_\text{pair}&=& V''_\text{pair}
  \sum_{\langle\langle i,k,j\rangle\rangle,\sigma}\Big[
  c^\dagger_{k,\sigma} c^\dagger_{k,\bar{\sigma}} 
  c^{\phantom{\dagger}}_{i,\bar{\sigma}}n^{\phantom{\dagger}}_{i,\sigma}
  c^{\phantom{\dagger}}_{j,\sigma}(1-n^{\phantom{\dagger}}_{j,\bar{\sigma}})
  \nonumber \\
  && +c^\dagger_{k,\sigma} c^\dagger_{k,\bar{\sigma}} 
  c^{\phantom{\dagger}}_{i,\bar{\sigma}}(1-n^{\phantom{\dagger}}_{i,\sigma})
  c^{\phantom{\dagger}}_{j,\sigma}n^{\phantom{\dagger}}_{j,\bar{\sigma}}
  +\text{h.c.}\Big]\ .
\end{eqnarray}
The operators ${H}'_\text{pair}$ and ${H}''_\text{pair}$ describe
processes where two electrons from sites $i$ and $j$ are transferred
to site $k$. They do not change the number of DOs since at one site 
(with local operator $c_{i,\sigma}(1-n_{i,\bar{\sigma}})$) a DO is
created while at another ($c_{i,\sigma}n_{i,\bar{\sigma}}$) a DO is
annihilated. The local operator $c^\dagger_{k,\uparrow}
c^\dagger_{k,\downarrow}$ turns an empty site into a doubly occupied
site and thus does not change the number of DOs in our counting.
Both the coefficients $V'_\text{pair}$ and
$V''_\text{pair}$ change only slightly on passing from calculation (iii) 
to calculation (iv); the difference between (iv) and (v) is minute so
that we consider the final result as reliable for not too large $W/U$
where the mapping to the generalized $t$-$J$ models is possible.

The last type of operators created in second order in $t/U$ are
correlated hopping terms. We classify them as interactions because they have a
non-vanishing effect only if at least two double occupancies or holes are 
present.  The operator 
\begin{eqnarray}
  V'_{n,0}&=& V'_n \sum_{\langle i,k,j\rangle; \alpha,\beta}
  \Big[
  (1-n^{\phantom{\dagger}}_{i,\alpha}) c^\dagger_{i,\overline{\alpha}}
  c^{\phantom{\dagger}}_{j,\overline{\beta}} 
  (1-n^{\phantom{\dagger}}_{j,\beta}) \bar{n}^{\phantom{\dagger}}_k
  \nonumber \\ 
  && \hspace*{1cm}
  + n^{\phantom{\dagger}}_{i,\alpha} c^\dagger_{i,\overline{\alpha}}
  c^{\phantom{\dagger}}_{j,\overline{\beta}}
  n^{\phantom{\dagger}}_{j,\beta}
  \bar{n}^{\phantom{\dagger}}_k +\text{h.c.}  \Big]
\end{eqnarray}
describes a process of a DO hopping between the NNN sites $i$ and $j$
 if and only if there is a DO on the site $k$. The
corresponding operator for a 3NN process is
\begin{eqnarray}
  V''_{n,0}&=& V''_n \sum_{\langle\langle i,k,j\rangle\rangle; \alpha,\beta}
  \Big[(1-n^{\phantom{\dagger}}_{i,\alpha}) c^\dagger_{i,\overline{\alpha}}
  c^{\phantom{\dagger}}_{j,\overline{\beta}} 
  (1-n^{\phantom{\dagger}}_{j,\beta})\bar{n}^{\phantom{\dagger}}_k
  \nonumber \\ 
  && \hspace*{1cm}
  + n^{\phantom{\dagger}}_{i,\alpha} c^\dagger_{i,\overline{\alpha}}
  c^{\phantom{\dagger}}_{j,\overline{\beta}}
  n^{\phantom{\dagger}}_{j,\beta} \bar{n}^{\phantom{\dagger}}_k
  +\text{h.c.}  \Big]. 
\end{eqnarray}
The coefficients $V'_n$ and $V''_n$ are shown in Fig.\ \ref{fig:V}d
and e. 

The above results define the charge dynamics of the $t$-$J$ model
in a systematic and exhaustive way. We highlight the importance of
spin-dependent hoppings.
All the values for the hopping coefficients and the interactions show
excellent convergence in the relevant region up to $W/U\approx 0.9$. 
We find again that for values $W/U > 0.9$ the coefficients display 
anomalous behavior. This effect is most striking for the coefficient
$V'$ in Fig.~\ref{fig:U}c. We attribute this spurious behavior
to the breakdown of the mapping of the Hubbard model to a generalized
$t$-$J$ model without specifying the state of the spin degrees
of freedom.

\section{Discussion}\label{sec:dis}
In this section, we will discuss the physical significance of our findings for
the square lattice and compare them to the situation for the Hubbard model in 
infinite dimensions.

In the derivation of the effective $t$-$J$-model, one does not aim at the 
solution of the complete problem of the Hubbard model on the square lattice.
The CUT is designed to disentangle the energetically 
low-lying dynamics, i.e., the dynamics of the spins and of the doped holes,
from processes involving more double occupancies. So no
particular state in spin space is assumed, but an effective
model without processes changing the number of double occupancies is derived.
In this calculation, the paramagnetic ensemble at half-filling serves as
reference ensemble.

The general consideration presented in the Introduction suggested
that the reduction to the effective model without
charge fluctuations cannot be defined for too large $W/U$. Our findings fully 
corroborate this view. The reduction of the Hubbard model to a
generalized $t$-$J$ model is possible as long as the apparent gap 
$\Delta_\text{app}$ remains non-negative. The appearance of negative values 
for $\Delta_\text{app}$
is an artefact. It indicates that the paramagnetic reference ensemble 
does no longer represent a phase which is stable against
fluctuations involving double occupancies. 
In this case the reduction to an effective model without
virtual double occupancies is no longer possible.

The concept of the apparent gap is introduced as a quantitative 
measure of the energetic separation of the low-lying degrees of
freedom from the fluctuations involving double occupancies. But it is not
easy to visualize since it does not refer to true eigen states of
the system. So it is helpful for a qualitative understanding to consider
a modified system where the apparent and the true gap coincide.
This situation is realized for a Hubbard model on the Bethe lattice
in the limit of infinite coordination number $Z\to \infty$, cf.\ Refs.\
\onlinecite{georg96,gebha97,eastw03,nishi04b}. For infinite
dimensional lattices it is assumed that long-range order can be completely
suppressed by frustration, e.g., in the generalizations of fcc lattices to
infinite dimensions \cite{mulle91,uhrig96a}.  Since the magnetic 
couplings $J$ scale with the inverse coordination number $Z^{-1}$ 
there are no  magnetic correlations at all once the static sublattice 
magnetization is suppressed \cite{mulle89a}. 
This is so since the second moment  of the magnetic couplings, averaged over 
adjacent sites, vanishes in the limit $Z\to \infty$. There is even no
nearest-neighbor correlation. Hence the paramagnetic reference ensemble 
represents the highly degenerate magnetic ground state in infinite dimensions
without long-range order. The apparent gap and the true charge
gap are identical. Their closure signals a real insulator-to-metal transition.
In finite dimensions the closure of the apparent gap signals
only a crossover from well-separated to non-separated energy scales.

We see that the (paramagnetic) insulator-to-metal \emph{crossover}
in finite dimensions can be studied in the purified form of a phase transition
in the infinite dimensional Hubbard model. Indeed, similar behavior
is found. The single-particle gap $\Delta$ vanishes in infinite dimensions
at about $1.1W -1.2 W$ \cite{bulla01a,garci04,blume04,nishi04b,eastw03}. 
It disappears roughly linearly as function of 
$U$ \cite{eastw03,nishi04b}. These similarities corroborate the infinite
dimensional model as illustration for the crossover in finite dimensions.

The validity of our approach is clear for small values of $W/U$. The
only approximation used is the truncation of processes of a spatial 
range $d\ge 4$. The remaining question is to know up to which ratio of
$W/U$ the results are reliable. Our findings in finite dimensions
illustrate that the answer to this question is difficult already on the 
conceptual level. The apparent gap $\Delta_\text{app}$ measuring the 
separation of energy scales does not compare the energies of real 
eigen states. Hence we claim that the validity of the mapping of the 
Hubbard model to a generalized $t$-$J$ model is not limited by a sharply
defined transition but by a gradual crossover. 

An estimate up to which ratio of $W/U$ the mapping is reasonable
can be obtained from the closure of the apparent gap at around 
$W/U \approx 0.9$. A better estimate
for the region of validity should compare $\Delta_\text{app}$ to the size of 
the coefficients of the neglected operators. But these coefficients are
not available. Hence we compare $\Delta_\text{app}$ to the size of the 
coefficients of the operators with the \emph{maximal} extension considered, 
that is $d=3$. Both energies become equal for $W/U\approx 0.85$. If in addition
we take the uncertainty of the extrapolation procedure into 
account we arrive at the conservative estimate that
the generalized $t$-$J$ model can be used up to $W/U\approx 0.8$
which includes the commonly assumed parameters for
cuprate planes $t/J\approx 3$ which translates to $W/U\approx 0.7$.
Up to $W/U\approx 0.8$, the quantitative results provided in this article 
are reliable.

In conclusion, we find that the possibility to map a Hubbard model
in a systematic and controlled way to a generalized
$t$-$J$ model depends essentially on local physics. We stress, however,
that the breakdown of this mapping does not represent a real physical
phase transition in finite dimensions. We suppose that the breakdown
of the mapping manifests itself in physical properties as a crossover.
It is plausible, for instance, that the nature of the elementary
excitations changes:
holes and magnetic modes for large values of the interaction become
fermionic quasi-particles at small values of the interaction.

\section{Summary}\label{sec:sum}
We introduced a self-similar CUT to map the Hubbard model at strong
coupling onto an effective model. 
The generator of the transformation is chosen such that 
the number of double occupancies is conserved by the effective
model which is thus a generalized $t$-$J$ model. The CUT allows
to do this in a systematic and non-perturbative way which
includes the dynamics of doped holes as well as the spin dynamics.
Thus our analysis extends previous ones in two important aspects:
the non-perturbative treatment of the couplings and the 
systematic treatment of the dynamics of doped charge carriers.

The calculations confirmed our expectations which were based on qualitative
arguments. The fundamental result is that for $U\gtrsim 1.3 W$ 
the mapping of the Hubbard model to a generalized $t$-$J$ model is possible. 
The parameters of  the effective model can be determined reliably by the CUT. 
They are governed  mainly by local physics. On the other hand, the reduction 
of the Hubbard model to an effective model without charge fluctuations is not 
possible for $U\lesssim 1.2 W$.  Here, the sectors of different number of 
double occupancies cannot be disentangled.

The truncation scheme necessary to close the flow equation of the CUT was based
on the extension of the operators. In order to define the
\emph{extension} of an operator a physically meaningful unique
notation is required. This was achieved by introducing a
normal-ordering with respect to a vacuum. The vacuum chosen here was
the paramagnetic ensemble without any magnetic correlations.

Besides the NN Heisenberg exchange further 
two-spin terms and four-spin terms were taken into account.  It was shown that 
two-spin couplings other than the Heisenberg term are small
whereas the four-spin ring exchange $J_\Box$
gives a sizeable contribution to the effective spin model. 
Results for the size of the coefficients for NN hopping, NNN hopping
and spin-dependent NNN hopping were obtained. It was found that the 
spin-dependent NNN and 3NN hopping is as important as the spin-independent one.
We recommend to account for this fact in phenomenological parametrizations
of experimental Fermi surfaces.
In addition, the size of the interactions between holes and double
occupancies on NN and NNN sites were calculated. Various truncation schemes
show very good convergence of the coefficients of the effective model
in the relevant parameter region $W\lesssim 0.8 U$. 

Our results suggest that self-similar continuous unitary transformations can
be used also for other models, e.g.\ more realistic multi-band Hubbard models, 
in order to achieve a systematic and quantitative reduction to effective 
models.

\acknowledgments
We thank K.P.~Schmidt for stimulating discussions and the DFG for
financial support in SFB 608.

\appendix
\section{The NN truncation}\label{app:NN}
To carry out the CUT with all operators on NN sites the terms
(\ref{minv}) and (\ref{minpair}) in equation (\ref{mingesamt}) have to be
included in the Hamiltonian
\begin{eqnarray}
  {H}_V &=& V(\ell) \sum_{\langle i,j\rangle} \bar{n}_i\bar{n}_j \\
  {H}_\text{pair}&=& V_\text{pair}(\ell) 
  \sum_{\langle i,j\rangle} c^\dagger_{i,\uparrow} c^\dagger_{i,\downarrow} 
  c^{\phantom{\dagger}}_{j,\downarrow} c^{\phantom{\dagger}}_{j,\uparrow}\ .
\end{eqnarray}
$H_V$ describes interactions of electrons on NN sites, $H_{\rm pair}$
the hopping of two electrons between site $i$ and $j$. These terms do
not change the number of DOs. Thus, they generate no 
new contributions in $\eta$. Calculating 
$[\eta,H_\text{NN}+ H_V+H_\text{pair}]$ and
neglecting again terms that do not fit on NN sites we obtain the 
closed flow equation
\begin{subequations}
\begin{eqnarray}
  \label{dglNN}
  d_\ell U(\ell)&=& 32 t_{+2}(\ell)^2 \\
  d_\ell t_{0}(\ell)&=& 0                      \\
  d_\ell t_{+2}(\ell)&=& -2 U(\ell)t_{+2}(\ell) \\
  &&+t_{+2}(\ell)\left( 2 V(\ell)-2 V_\text{pair}(\ell)-3
  J(\ell)/2)\right) \nonumber \\
  d_\ell J_1(\ell)&=& 16 t_{+2}(\ell)^2 \\
  d_\ell V(\ell)&=& -4 t_{+2}(\ell)^2 \\
  d_\ell V_\text{pair}(\ell)&=& 8 t_{+2}(\ell)^2\ . 
\end{eqnarray}
\end{subequations}
The solution for $J_1$ is
\begin{eqnarray}
  \label{solNN}  
J_1^{\rm NN}(\ell)= \frac{2r}{7}  \tanh(2r \ell +D )-\frac{2}{7} U_0 
\end{eqnarray}
where $r={\sqrt{28 t_0^2 + {U_0}^2}}$ and $D=\mathrm{arctanh}(U_0/r)$.
This yields in the limit $\ell \to \infty$ the effective
NN spin-coupling of the NN truncation
\begin{equation}
  \label{J1NN}  
  J_{1,{\rm eff}}^{\rm NN}= \frac{2}{7}
     {\sqrt{U^2+28 t^2}} -\frac{2}{7}U\ ,
\end{equation}
where the final result is given in the bare parameters
recalling $t_0=t, U_0=U$.


\begin{thebibliography}{10}
\bibitem{notiz4}
The Hamiltonian (\ref{hamilton}) changes the
number of DOs only by a single absolute value, namely $2$. So the only
difference between the matrix elements $\text{sign}(d_i - d_j)H_{ij}$ and 
$(d_i - d_j)H_{ij}$ in an eigen basis of 
$\hat D$ is a factor of 2.
 .
\bibitem{notiz3}
We use the expression
``normal-ordering'' while we are aware that we are not dealing
with conventional normal-ordering since the strong-coupling problem
at hand does not have non-interacting fermions or bosons.
Normal-ordering in the context of the present work refers to 
the standardized procedure to represent operators.
\bibitem{notiz2}
Takahashi gives the 4th order result in Ref.~\onlinecite{takah77}. The
calculation of the 6th order was done using the method described by
Takahashi\cite{takah77}. 
.
\end{thebibliography}

\end{document}